\documentclass[aip,pof,amsmath,amssymb,preprint]{revtex4-1}

\usepackage{tabularx}
\newcolumntype{C}[1]{>{\centering\arraybackslash}p{#1}}
\usepackage{textcase}
\usepackage{graphicx}
\usepackage{units}
\usepackage{siunitx}
\usepackage{color}
\usepackage{bm}
\usepackage{multirow}
\usepackage{tikz}
\usetikzlibrary{shapes,arrows}
\usepackage{pgfplots}
\usepgfplotslibrary{groupplots}
\usetikzlibrary{spy,decorations.fractals,shapes.misc}
\usetikzlibrary{decorations.markings}
\usepgfplotslibrary{colormaps}

\usepackage[caption=false]{subfig}
\usepackage{epstopdf}

\begin{document}

\title[Comparison of DSMC and particle based BGK models]{Particle-based fluid dynamics: comparison of different Bhatnagar-Gross-Krook models and the Direct Simulation Monte Carlo method for hypersonic flows} 

\author{M. Pfeiffer}
 \email{mpfeiffer@irs.uni-stuttgart.de}
\affiliation{%
Institute of Space Systems, University of Stuttgart, Pfaffenwaldring 29, D-70569
   Stuttgart, Germany
}%

\date{\today}

\begin{abstract}
The Bhatnagar-Gross-Krook (BGK) model as well as its extensions (ellipsoidal statistical BGK, Shakhov BGK, unified BGK)
are used in particle-based fluid dynamics and compared with the Direct Simulation Monte Carlo (DSMC) method. 
To this end, different methods are investigated that allow efficient 
sampling of the Shakhov and the unified target distribution functions.
As a consequence, particle simulations based on the Shakhov BGK and the unified BGK model are possible in an efficient way.
Furthermore, different energy conservation schemes are tested for the BGK models. It is shown that the choice
of the energy conservation scheme has a major impact on the quality of the results for each model.
The models are verified with a Couette flow problem at different Knudsen numbers and wall velocities. Furthermore, the models are compared with DSMC results of a hypersonic
flow around a 70$^\circ$ blunted cone. 
It is shown that the unified BGK model is able to reproduce rarefied gas phenomena. Furthermore, it is shown that the difference in 
the reproduction of shock structure is not significant between the ellipsoidal statistical BGK and Shakhov BGK model for the 
flow around a 70$^\circ$ blunted cone and significantly depends on the energy conservation method. The choice of the energy conservation method is especially crucial for 
the Shakhov model whereas the ellipsoidal statistical BGK model is much more robust concerning the energy conservation scheme.
Additionally, a computational
time study is done to show the efficiency of BGK-based simulations for low Knudsen number flows compared with DSMC.
\end{abstract}


\maketitle

\section{Introduction}
Simulations of gas flows which cover a wide range of Knudsen numbers including continuum and rarefied gas regions are still challenging.
CFD methods based on Navier-Stokes equations cover a wide range of near equilibrium flows that are important for many practical applications. Nevertheless, the assumptions
behind the Navier-Stokes equations become invalid for rarefied non-equilibrium flows. For this purpose, 
the Direct Simulation Monte Carlo (DSMC) method can be used to treat non-equilibrium effects in rarefied gases \cite[]{Bird1994}.
Therefore, the collision integral is not solved directly. Consequently, the DSMC method becomes very expensive for small Knudsen number flows due to the fact 
that molecular events must be spatially- and time-resolved within the mean free path and collision frequency scales, respectively.
To close the gap between the applicable regimes of both methods, different approaches are used. 

A straightforward approach to overcome this gap is the coupling of DSMC with CFD. Here, many problems arise due to the very
different underlying approaches of both methods. Especially the statistical noise of the DSMC method is problematic at the boundaries
between DSMC and CFD. Another possibility are particle-based continuum methods, which can be easily
coupled with DSMC.
A variety of different methods were introduced in the past, e.g. the Time Relaxed Monte Carlo method \cite[]{Pareschi2005}, 
the Viscous Collision Limiter method \cite[]{Liechty2016}, the Fokker-Planck solution algorithm \cite[]{Gorji2014,PFEIFFER20171}, 
the low diffusion method \cite[]{Burt2008,MIRZA2017269} and more. A short overview with advantages and disadvantages of these methods 
is given in Mirza et al. \cite{MIRZA2017269}.

Another particle method that is already used and coupled to DSMC in different applications like nozzle flow expansion \cite[]{burt2006evaluation}, 
micro channel flows (MEMS) simulations \cite[]{titov2008analysis} or simulations of
hypersonic shocks \cite[]{tumuklu2016particle} is the statistical Bhatnagar-Gross-Krook (BGK) method. 
Up to now, only the ellipsoidal statistical BGK (ESBGK) model has been
used in this context. Unfortunately, the ESBGK model has difficulties to reproduce the shock structure correctly \cite[]{chen2015comparison}. In this publication, 
methods to sample in an efficient way from other target distribution functions used in the Shakhov and unified BGK models are described. Furthermore, different energy and momentum conservation
schemes are investigated, which are especially critical for the Shakhov model to achieve a correct physical behavior. 
Finally, different BGK models 
(BGK, ESBGK, Shakhov BGK, Unified BGK) will be verified by simulating Couette flow problems and tested as well as compared concerning the reproduction of hypersonic shocks.

\section{Theory}
The Boltzmann equation fully describes the behavior of a monoatomic gas with the corresponding distribution function 
$f=f(\mathbf x, \mathbf v, t)$ at position $\mathbf x$ and velocity $\mathbf v$
\begin{equation}
\frac{\partial f}{\partial t} + \mathbf v \frac{\partial f}{\partial \mathbf x} = \left.\frac{\partial f}{\partial t}\right|_{Coll}.
\end{equation}
In this equation, external forces are neglected. Furthermore, $\left.\partial f/\partial t\right|_{Coll}$ is the 
collision term, which can be described by the Boltzmann collision integral
\begin{equation}
\left.\frac{\partial f}{\partial t}\right|_{Coll}=\int_{\mathbb{R}^3}\int_{S^2}\mathcal B
\left[f(\mathbf v')f(\mathbf v_*')-f(\mathbf v)f(\mathbf v_*)\right]d\mathbf n d\mathbf v_*.
\end{equation}
Here, $S^2\subset\mathbb{R}^3$ is the unit sphere, $\mathbf n$ is the unit vector of the scattered velocities, $\mathcal B$ is the collision
kernel and the superscript $'$ denotes the post collision velocities. The multiple integration of this collision term makes it difficult to compute in the 6D phase space.

\subsection{BGK Models}

The BGK model approximates the collision term to a simple relaxation form where the distribution function relaxes towards a target distribution function $f^t$ with a 
certain relaxation frequency $\nu$:
\begin{equation}
\left.\frac{\partial f}{\partial t}\right|_{Coll}=\nu\left(f^t-f\right).
\label{eq:bgkmain}
\end{equation}
The original BGK model assumes, that the target velocity distribution function is the Maxwellian velocity distribution $f^M$
\begin{equation}
f^M=n\left(\frac{m}{2\pi k_B T}\right)^{3/2} \exp\left[-\frac{m\mathbf c^2}{2k_B T}\right],
\label{eq:maxwelldist}
\end{equation}
with the particle density $n$, particle mass $m$, temperature $T$, Boltzmann constant $k_B$ and the thermal particle velocity $\mathbf c=\mathbf v -\mathbf u$ from the 
particle velocity $\mathbf v$ and the average flow velocity $\mathbf u$ \cite[]{bhatnagar1954model}. The BGK model reproduces the Maxwellian distribution in an
equilibrium state, preserves conservation of mass, momentum and energy as well as fulfills the H-theorem \cite[]{gallis2011investigation}. 
The relaxation frequency
$\nu$ defines the viscosity $\mu$ and the thermal conductivity $K$
\begin{equation}
\mu=\frac{nk_BT}{\nu} \qquad\qquad K=\frac{c_Pnk_BT}{\nu}
\end{equation}
with the specific heat constant 
$c_P=5k_B/2m$. These equations clarify the problem of the BGK model concerning the Prandtl number, which is fixed in this model to 
$Pr=\mu c_P/K=1$, whereas the Prandtl number of monoatomic gases is $Pr\approx 2/3$ \cite[]{vincenti1965introduction}. As a consequence, only the 
thermal conductivity or viscosity can be correctly reproduced at a time with the BGK model. To overcome this problem,
several extensions of the BGK model were introduced in the past. Some of these models transform the target distribution function
e.g. the ellipsoidal statistical BGK model \cite[]{holway1966new} or the Shakhov BGK model \cite[]{shakhov1968generalization}, 
other models change the relaxation frequency from a constant to a function of the microsopic velocities as described in Struchtrup\cite{struchtrup1997bgk}. 
In this paper, only models with a transformed target distribution are investigated.

\subsubsection{Ellipsoidal Statistical BGK Model}
The ellipsoidal statistical BGK (ESBGK) model replaces the Maxwellian target distribution of the standard BGK model with an 
anisotropic Gaussian distribution \cite[]{gallis2011investigation}
\begin{equation}
f^{ES}=\frac{n}{\sqrt{\det \mathcal A}} \left(\frac{m}{2\pi k_B T}\right)^{3/2} \exp\left[-\frac{m\mathbf c^T \mathcal A^{-1} \mathbf c}{2k_B T}\right]
\label{eq:esbgkdist}
\end{equation}  
with the anisotropic matrix 
\begin{equation}
\mathcal A = \mathcal I - \frac{1-Pr}{Pr}\left(\frac{3\mathcal P}{\mathrm{Tr}\left[\mathcal P\right]}-\mathcal I\right).
\end{equation}
The anisotropic matrix $\mathcal A$ consists of the identity matrix $\mathcal I$ and the pressure tensor $\mathcal P$
\begin{equation}
\mathcal P = \int \mathbf c \mathbf c^T f\,d\mathbf v,
\end{equation}
which are both symmetric. The ESBGK model reproduces the Maxwellian distribution in the equilibrium state as well as the correct moments
of the Boltzmann equation. Furthermore, Andries et al.\cite{andries2000gaussian,andries2001bgk} have shown that it
fulfills the H-theorem. In the ESBGK model, the viscosity and the thermal conductivity are defined as 
\begin{equation}
\mu=\frac{nk_BT}{\nu}Pr \qquad\qquad K=\frac{c_Pnk_BT}{\nu}.
\end{equation}
Due to the fact that the viscosity depends on the Prandtl number, it is now possible to reproduce the viscosity and thermal
conductivity at the same time. So, the introduction of the Prandtl number as an additional parameter resolves the Prandtl number
problem of the standard BGK model.

As proposed by Gallis and Torczynski\cite{gallis2011investigation}, a symmetric transformation matrix $\mathcal S$ with $\mathcal A = \mathcal S \mathcal S$ can be defined. 
Furthermore, a normalized thermal velocity vector $\mathbf C$ is defined such that $\mathbf c= \mathcal S \mathbf C$. 
Using these definitions, the argument of the exponential function in Eq. \eqref{eq:esbgkdist} becomes
\begin{equation}
\mathbf c^T \mathcal A^{-1} \mathbf c = (\mathcal S\mathbf C)^T \mathcal S^{-1} \mathcal S^{-1} \mathcal S\mathbf C = \mathbf C^T \mathbf C  
\label{eq:smat}
\end{equation}
using $(\mathcal S\mathbf C)^T=\mathbf C^T \mathcal S^T=\mathbf C^T \mathcal S$ due to the fact that $\mathcal S$ is symmetric. So,
$\mathcal S$ can transform a vector $\mathbf C$ sampled from a Maxwellian distribution to a vector $\mathbf c$ sampled from Eq.
\eqref{eq:esbgkdist}.

\subsubsection{Shakhov BGK Model}
In contrast to the ESBGK model, which modifies the shear stress to produce the correct Prandtl number \cite[]{kun2014direct}, the Shakhov model
(SBGK) directly modifies the heat flux. For this, the target distribution of the BGK model is changed to 
\begin{equation}
f^S=f^M\left[1+(1-Pr)\frac{\mathbf c \mathbf q}{5n(RT)^2}\left(\frac{\mathbf c^2}{2RT}-\frac{5}{2}\right)\right]
\end{equation}
with the heat flux vector
\begin{equation}
\mathbf q = \int \mathbf c \mathbf c^2 f\,d\mathbf v,
\label{eq:heatflux}
\end{equation}
and the specific gas constant $R$.
In the SBGK model, the viscosity and the thermal conductivity are defined as 
\begin{equation}
\mu=\frac{nk_BT}{\nu} \qquad\qquad K=\frac{c_Pnk_BT}{\nu}\frac{1}{Pr}.
\end{equation}
Consequently, this model solves the Prandtl number problem of the original BGK model and reproduces the Maxwellian distribution 
in the equilibrium state as well as the correct moments of the Boltzmann equation. The main disadvantage of 
the SBGK model compared to the ESBGK model is that no general proof exists whether the SBGK model always fulfills the H-theorem.
Furthermore, it cannot be guaranteed that $f^S$ is positive in each situation \cite[]{kun2014direct}.

\subsubsection{Unified BGK Model}
\label{sec:ubgk}
Chen et al.\cite{chen2015comparison} proposed a unified BGK model (UBGK), which merges the ESBGK and SBGK model. The new target distribution
is 
\begin{eqnarray}
f^U &=& f^{ES*}+f^{S*}\\
f^{ES*} & =&\frac{n}{\sqrt{\det \mathcal A^*}} \left(\frac{m}{2\pi k_B T}\right)^{3/2} \exp\left[-\frac{m\mathbf c^T \mathcal A^{*-1} \mathbf c}{2k_B T}\right]\\
\mathcal A^* &=&\mathcal I + C_{ES}\left(\frac{3\mathcal P}{\mathrm{Tr}\left[\mathcal P\right]}-\mathcal I\right)\\
f^{S*} &=& f^M(1-C_S)\frac{\mathbf c \mathbf q}{5n(RT)^2}\left(\frac{\mathbf c^2}{2RT}-\frac{5}{2}\right).
\end{eqnarray}
The UBGK model depends on two different parameters $C_{ES}$ and $C_S$ with 
\begin{equation}
Pr=\frac{C_S}{1-C_{ES}}.
\end{equation}
Therefore, this model introduces an additional independent parameter for a derived Prandtl number. This parameter can basically be 
chosen freely but the ES model constrains $C_{ES}$ to the interval $[-0.5,1)$ to preserve positive eigenvalues of $\mathcal A$ \cite[]{chen2015comparison}. 
When $C_{ES}=0$ and $C_S=Pr$ the UBGK model reduces to the Shakhov model. On the other hand, if $C_{ES}=1-1/Pr$ and
$C_S=1$ the UBGK model is identical to the ESBGK model. For other values, the UBGK model blends the ESBGK and Shakhov
model. The viscosity and the thermal conductivity for the UBGK model are defined as 
\begin{equation}
\mu=\frac{nk_BT}{\nu(1-C_{ES})} \qquad\qquad K=\frac{c_Pnk_BT}{\nu}\frac{1}{C_S}.
\end{equation}
Finally, the UBGK model allows to modify the shear stress as well as the heat flux at once. Or with other words, the parameters
$C_{ES}$ and $C_S$ can be used to determine different relaxation rates of $\mathcal P$ and $\mathbf q$ \cite[]{kun2014direct}:
\begin{equation}
\frac{\partial \mathcal P}{\partial t}=\frac{-(1-C_{ES})}{\nu}\mathcal P, \qquad
\frac{\partial \mathbf q}{\partial t}=\frac{-C_{S}}{\nu}\mathbf q.
\end{equation}
The problem of the UBGK model is that it is not clear how $C_{ES}$ and $C_S$ should be chosen. Chen et al.\cite{chen2015comparison} exemplary show that
when $C_S$ is used to produce the right Prandtl number, $C_{ES}$ can be derived as
\begin{equation}
C_{ES}^{VHS}=1-\frac{(7-2\omega)(5-2\omega)}{30},
\label{eq:cesvhs}
\end{equation}
for VHS molecules with the VHS parameter $\omega$ in an equilibrium state. 
However, they also showed that $C_{ES}$ totally differs for non-equilibrium states.

\section{Implementation}
The different BGK models are implemented in the PIC-DSMC code PICLas \cite[]{Munz2014} and verified by the comparison to DSMC results of PICLas. 
The DSMC method is widely used and details can be found in Bird\cite{Bird1994}. 

For the BGK models, the statistic particle method of Gallis and Torczynski\cite{gallis2011investigation,gallis2000application} is used. 
Here, particles are moved in a simulation mesh,
collide with boundaries and microsopic particle properties are sampled to calculate macroscopic values in the same manner as in 
DSMC. But in contrast to the DSMC method, the collision step with binary collisions between the particles is not performed. Instead,
each particle in a cell relaxes with the probability
\begin{equation}
P=1-\exp\left[-\nu \Delta t\right]
\end{equation}
according to Eq. \eqref{eq:bgkmain} towards the target distribution. The relaxation frequency $\nu$ is evaluated in each time step for each
cell from the definition of the viscosity of each model. The relaxation frequency directly depends on the cell temperature $T$, which
is calculated from the particle information. For the viscosity $\mu$ the well known exponential ansatz
\begin{equation}
\mu=\mu_{ref}\left(\frac{T}{T_{ref}}\right)^\omega
\end{equation}
is used. Here, $T_{ref}$ is a reference temperature and $\mu_{ref}$ the reference dynamic viscosity at $T_{ref}$ \cite[]{burt2006evaluation}. 
For a VHS gas the reference dynamic viscosity can be calculated with the VHS reference diameter $d_{ref}$ of the particles:
\begin{equation}
\mu_{ref}=\frac{30\sqrt{mk_BT_{ref}}}{\sqrt{\pi}4(5-2\omega)(7-2\omega)d_{ref}^2}.
\end{equation}

If a particle is chosen to relax, the new particle velocity is sampled from the target distribution.

\subsection{Energy and Momentum Conservation}
\label{sec:encon}
To conserve momentum and energy, two different methods are proposed in the literature. 
In the mehtod of Gallis and Torczynski\cite{gallis2011investigation}, the average velocity and the temperature for each particle $N$ in a cell 
are determined before the collision ($\mathbf u$ and $T$) and for the provisional post collision conditions ($\mathbf u_p$ and $T_p$):
\begin{eqnarray}
\mathbf u &= \frac{1}{N} \sum_{i=1}^N \mathbf v_i \\
T &= \frac{1}{N-1} \sum_{i=1}^N \frac{m \mathbf c_i^2}{3 k_B}.
\end{eqnarray}
The factor $\frac{1}{N-1}$ in the temperature occurs to produce unbiasedness of $T$\cite{sun2005evaluation}.
The final
postcollision velocity $\mathbf v^*$ of every molecule (whether having undergone relaxation or not) is then determined from the 
provisional postcollision velocity $\mathbf v_p$ according to
\begin{equation}
\mathbf v^*=\mathbf u +(\mathbf v_p- \mathbf u_p)\sqrt{\frac{T}{T_p}}.
\end{equation}

In the method of Burt and Boyd\cite{burt2006evaluation}, the same guarantee is performed but only with $N_{relax}$ particles that take part in the 
relaxation process whereas all other particles will remain unaffected. Obviously, this method can only be perfomed if at least two particles
are chosen to relax. To handle the case $N_{relax}=1$, a random number is used to decide whether $N_{relax}=0$ or $N_{relax}=2$ with equal probability. 
If $N_{relax}$ is determined to two, the second particle is chosen randomly from the remaining not relaxing particles.

\subsection{Relaxation Process}
In the standard BGK collision term, the postcollision velocities $\mathbf v_p$ are simply sampled from the Maxwellian distribution
\eqref{eq:maxwelldist}
\begin{equation}
\mathbf v_p = \bm{\xi} \sigma_{T},
\end{equation}
with the normal distributed random vector $\bm{\xi}$
and the velocity $\sigma_{T}=\sqrt{k_BT/m}$.

\subsubsection{Sampling from ESBGK}
\label{sec:esbgkimpl}
The sampling from the ESBGK distribution in Eq. \eqref{eq:esbgkdist} is performed with three different approaches. 

In the first approach 
an approximation of the transformation matrix $\mathcal S$ of Eq. \eqref{eq:smat} is used as described 
in previous studies \cite{gallis2011investigation, burt2006evaluation,tumuklu2016particle}
\begin{equation}
\mathcal S_{ij} = \delta_{ij}-\frac{1-Pr}{2Pr}\left[\frac{m}{k_B T}\frac{N}{N-1}\left(\mathcal P_{ij}-{\hat{c}_i}{\hat{c}_j}\right)-\delta_{ij}\right]
\label{eq:Sconvert}
\end{equation}
with 
\begin{equation}
\mathbf{\hat{c}} = \int \mathbf c f\,d\mathbf v.
\end{equation}
The advantage of this approach is that this method is fast and simple to implement. Nevertheless, the accuracy and performance of this
approximation has to be tested.

The second approach is the exact calculation of $\mathcal S$ by solving $\mathcal A = \mathcal S \mathcal S$ to find the square root of $\mathcal A$.
For this, the matrix $\mathcal A$ is diagonalized to the form
\begin{equation}
\mathcal A = \mathcal V \mathcal D \mathcal V^T.
\end{equation}
The columns of $ \mathcal V$ are the eigenvectors of $\mathcal A$ and the diagonal elements of the diagonal matrix $\mathcal D$ are the 
corresponding eigenvalues. The square root of $\mathcal A$ can now be calculated with
\begin{equation}
\mathcal S = \mathcal V \mathcal D^{1/2} \mathcal V^T,
\end{equation}
where the diagonal elements of the diagonal matrix $\mathcal D^{1/2}$ are the square roots of the eigenvalues.

\subsubsection*{Metropolis-Hastings Sampling}

\begin{figure}
\centering
\includegraphics{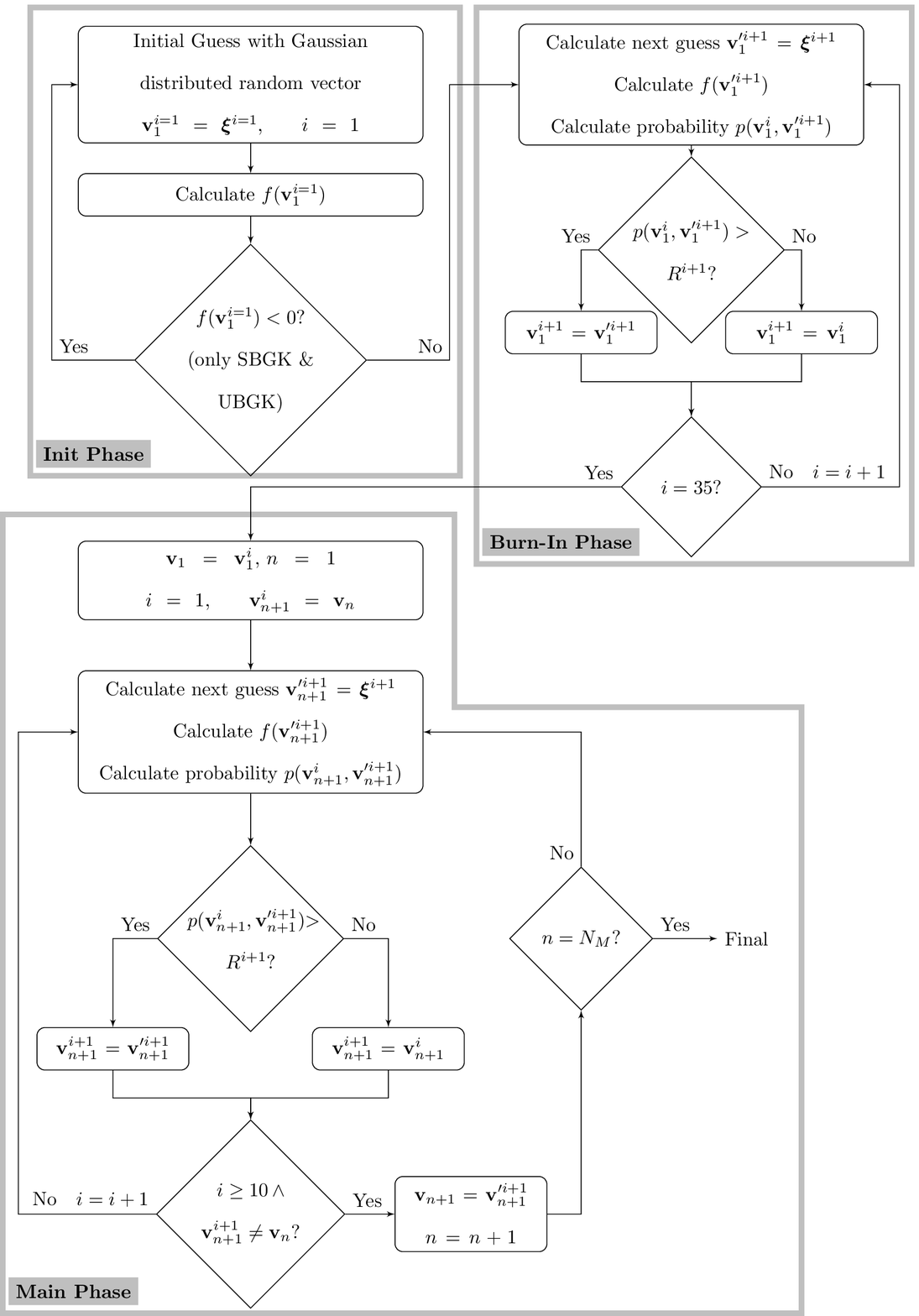}
\caption{Flowchart of the used Metropolis-Hastings method.}
\label{fig:mhflowchart}
\end{figure}

The third approach is the direct sampling from $f^{ES}$ by using a Metropolis-Hastings method 
which produces a Markov chain $\mathbf v_0$, $\mathbf v_1$, ..., $\mathbf v_{N_\mathrm{M}}$ of samples from the target distribution \cite[]{Chib1995, HASTINGS01041970, pfeiffer2016direct, garcia1998generation}. 
The MH method has the advantage that the determination of the global maximum of the target distribution function, which can be computationally expensive 
(e.g. for the proposed SBGK and UBGK distribution function), or the definition of an envelope function is not necessary in contrast to the acceptance-rejection (AR) method. 

Starting with a sample $\mathbf v_n$, a proposed sample $\mathbf v'_{n+1}$ is generated using a pre-specified density $q(\mathbf v_n,\mathbf v'_{n+1})$ and accepted as the next sample
$\mathbf v_{n+1}$ of the Markov chain with the probability 
\begin{equation}
p(\mathbf v_n,\mathbf v'_{n+1})=
\begin{cases}
\min\left(1,\frac{f(\mathbf v'_{n+1})q(\mathbf v'_{n+1},\mathbf v_n)}{f(\mathbf v_n)q(\mathbf v_n,\mathbf v'_{n+1})}\right),\qquad &f(\mathbf v_n)q(\mathbf v_n,\mathbf v'_{n+1})>0\\
1,\qquad &f(\mathbf v_n)q(\mathbf v_n,\mathbf v'_{n+1})=0.
\end{cases}
\label{eq:probmetrhast}
\end{equation}
If the proposed value $\mathbf v'_{n+1}$ is not accepted, the next step of the Markov chain is set to $\mathbf v_{n+1}=\mathbf v_n$.
A special case of the Metropolis-Hastings method (MH) is the random-walk Metropolis algorithm. Here, $q$ is a symmetrical function about zero, 
$q(\mathbf v_n,\mathbf v'_{n+1})=q(\mathbf v'_{n+1},\mathbf v_n)$ and the probability in Eq. \eqref{eq:probmetrhast} simplifies to
\begin{equation}
p(\mathbf v_n,\mathbf v'_{n+1})=\min\left(1,\frac{f(\mathbf v'_{n+1})}{f(\mathbf v_n)}\right).
\label{eq:metrprob}
\end{equation}
For the sampling from $f^{ES}$, $q$ is chosen as a Gaussian distribution, which is symmetric.
The detailed flowchart of the used MH algorithm is shown in Fig. \ref{fig:mhflowchart}.
First, the initial phase is used to find a proper start value of the MH algorithm. The start value is chosen with
\begin{equation}
\mathbf v_{1} = \bm{\xi}
\end{equation}
and the normal distributed random vector $\bm{\xi}$. 
Next, the distribution function $f(\mathbf v_{1})$ is calculated to check whether  
$f(\mathbf v_{1})$ is positive. This step is not necessary for the ESBGK distribution function but 
if the used target distribution function does not fulfill the positivity requirements for probabilities (e.g. the case for SBGK and UBGK), these samples must be discarded.
Therefore, another normal distributed random vector $\bm{\xi}$ is used until $f(\mathbf v_{1})>0$. A positive effect of the described MH algorithm is that if the first sample 
$\mathbf v_{1}$ fulfills $f(\mathbf v_{1})>0$, then also all following samples will fulfill this condition. Due to the fact that $p(\mathbf v_n,\mathbf v'_{n+1})$ would have to become
negative, such samples cannot be chosen anymore.

The next proposed value is detemined using again a new normal distributed random vector $\bm{\xi}$:
\begin{equation}
\mathbf v'_{n+1}=\bm{\xi}
\end{equation}
A characteristic of the MH algorithm is the so called \textit{burn-in} phase, which means that the first samples may not necessarily follow the target distribution, 
especially if the starting point is in a region of low density. Therefore, samples made during the \textit{burn-in} phase have to be discarded. 
Different simulations have shown that 35 initial steps are sufficient to overcome the \textit{burn-in} phase, before the first velocity is accepted. For each following
final accepted velocity in the main phase, the velocity should be changed at least one time and at least 10 steps in the Markov chain should be taken. Note, that the \textit{burn-in} phase
must be done in each cell every time step due to the changed target distribution.

For the ESBGK method, the probability $p(\mathbf v_n,\mathbf v'_{n+1})$ is 
\begin{equation}
p^{ESBGK}(\mathbf v_n,\mathbf v'_{n+1})=\min\left(1,\frac{\exp\left[-0.5\mathbf v^{'T}_{n+1} \mathcal A^{-1} \mathbf v'_{n+1}\right]}
{\exp\left[-0.5\mathbf v_{n}^T \mathcal A^{-1} \mathbf v_{n}\right]}\right)
\end{equation}
The final postcollision velocities are obtained by scaling the sampled velocities
\begin{equation}
\mathbf v_{n,p} = \mathbf v_{n} \sigma_{T}.
\label{eq:finalsamp}
\end{equation}

\subsubsection{Sampling from SBGK}
In contrast to the ESBGK model, an analytical expression to convert a normal distributed vector in a SBGK
distributed vector is not available (see Eq. \eqref{eq:Sconvert}). 
Therefore, the SBGK distribution is sampled using two different methods. The first method is the already mentioned MH algorithm.

For this, almost the same algorithm as for the ESBGK MH sampling is used, including the start and proposed velocities, the \textit{burn-in}
phase as well as the final scaling of the velocities (see Sec. \ref{sec:encon} for energy conservation details). The probability in Eq. \eqref{eq:metrprob} is changed to 
\begin{equation}
p^{SBGK}(\mathbf v_n,\mathbf v'_{n+1})=\min\left(1,\frac{\left(1+(1-Pr)\frac{\mathbf v'_{n+1}\mathbf q}{5n\sigma_T^{3/2}}
\left(\frac{\mathbf v^{'2}_{n+1}}{2}-\frac{5}{2}\right)\right)}
{\left(1+(1-Pr)\frac{\mathbf v_{n}\mathbf q}{5n\sigma_T^{3/2}}
\left(\frac{\mathbf v^2_{n}}{2}-\frac{5}{2}\right)\right)}\right).
\end{equation}
Note that unbiasedness of the heat flux vector from Eq. \eqref{eq:heatflux} can be achieved using
\begin{equation}
\mathbf q=n\frac{N}{(N-1)(N-2)}\sum_{i=1}^N \mathbf c_i \mathbf c_i^2.
\end{equation}

Instead of the MH method also the acceptance-rejection method can be used to sample from the Shakhov model. As shown for the Chapman-Enskog distribution in Garica and Alder\cite{garcia1998generation},
an envelope function $g(\bm{\xi})$ must be defined with $g(\bm{\xi})\ge f^S(\bm{\xi})$ for all $\bm{\xi}$. A detailed description how $g(\bm{\xi})$ should be defined can be 
found in Garica and Alder\cite{garcia1998generation}. However, good results for the following test cases are found for $g(\bm{\xi})=Af^M(\bm{\xi})$ with
\begin{equation}
A=1+4\max\left(\frac{\left| q_j \right|}{\sigma_T^{3/2}}\right),\quad j=1,2,3.
\end{equation}
The single steps for the acceptance-rejection sampling are:
\begin{enumerate}
\item Calculate $A$
\item Draw a normal distributed $ \mathbf v_{n} = \bm{\xi}$
\item Accept $ \mathbf v_{n} $ if random number $R<p^{SBGK,\,AR}(\mathbf v_{n})/A$
\item Calculate $v_{n,p} = \mathbf v_{n} \sigma_{T}$
\end{enumerate}
with $p^{SBGK,\,AR}(\mathbf v_{n})=\left(1+(1-Pr)\frac{\mathbf v_{n}\mathbf q}{5n\sigma_T^{3/2}}
\left(\frac{\mathbf v^{2}_{n}}{2}-\frac{5}{2}\right)\right)$.

\subsubsection{Sampling from UBGK}
The UBGK distribution is again sampled using the MH algorithm. As in the case of the SBGK sampling, only the probability in Eq. \eqref{eq:metrprob} is changed
\begin{eqnarray}
p^{UBGK}(\mathbf v_n,\mathbf v'_{n+1}) &=& \min\left(1,\frac{\hat{p}^{UBGK,1}(\mathbf v'_{n+1}) + \hat{p}^{UBGK,2}(\mathbf v'_{n+1})} 
{\hat{p}^{UBGK,1}(\mathbf v_{n}) + \hat{p}^{UBGK,2}(\mathbf v_{n})}\right), \\
\hat{p}^{UBGK,1}(\mathbf x)&=& \exp\left[-0.5\mathbf x^T \mathcal A^{-1} \mathbf x\right], \\
\hat{p}^{UBGK,2}(\mathbf x)&=& \exp\left[-0.5\mathbf x^2\right]\left((1-Pr)\frac{\mathbf x\mathbf q}{5n\sigma_T^{3/2}}
\left(\frac{\mathbf x^2}{2}-\frac{5}{2}\right)\right).
\end{eqnarray}
The UBGK method is not sampled using the acceptance-rejection method, due to the fact that the definition of the envelope function is cumbersome.

Note that the Metropolis-Hastings algorithm described here can be basically used to sample efficiently from an arbitrary target distribution 
and is not limited to the ESBGK, SBGK and the UBGK target distributions.

\section{Simulation Results}
\subsection{Couette Flow}
\label{sec:couette}
The first test case is a steady planar Couette flow of Argon similar to Kumar et al.\cite{kumar2009comparison}. For this, two parallel plates form a channel and move in opposite directions with a velocity 
$v_{Wall}$. All other boundaries are periodic, resulting in a 1D test case. The distance between the plates is $D=1\,\mathrm{m}$. The simulations are performed with 100 grid cells between the plates.
The DSMC solver in PICLas additionally uses an octree method \cite[]{Pfeiffer2013}. Therefore, each cell is subdivided to resolve the mean free path of the particles. The BGK methods are performed 
on the same mesh but without a mesh adaptation in this test case.

The used Argon parameters are $d_{ref}=4.17\cdot 10^{-10}\,\mathrm{m}$, $T_{ref}=273\,\mathrm{K}$, $\omega_{VHS}=0.81$ and $m=6.631368\cdot 10^{-26}\,\mathrm{kg}$. The initial gas temperature
as well as the temperature of the plates is in each simulation $T_{Wall}=273\,\mathrm{K}$. Additionally, diffuse reflection with full accomodation is assumed on the plates.

\subsubsection{Low Wall Velocity Case - $Kn\approx 0.014$}
\begin{figure}
\centering
\subfloat[Temperature\label{fig:lowvelo}]{\includegraphics{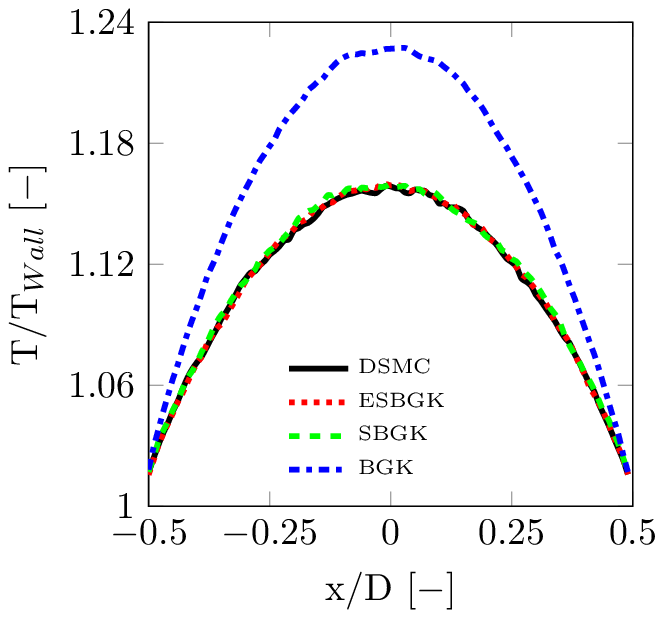}}\quad
\subfloat[Velocity\label{fig:lowvelob}]{\includegraphics{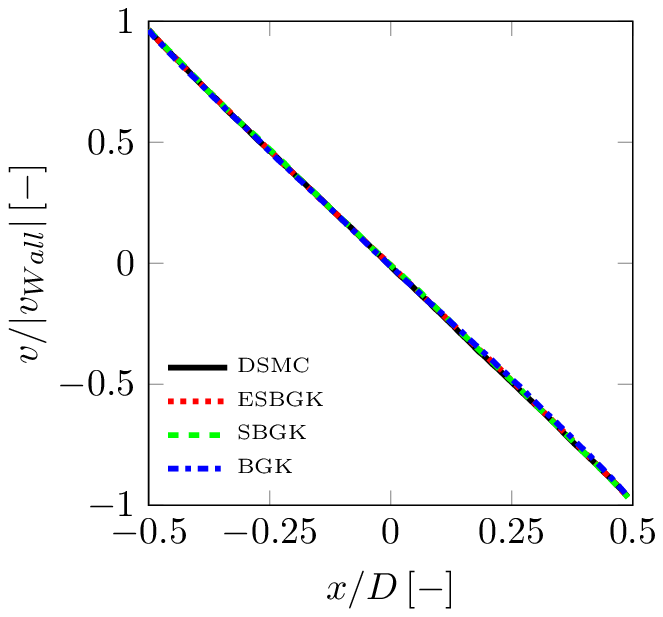}}\\
\subfloat[Comparison of sampling methods.\label{fig:lowveloc}]{\includegraphics{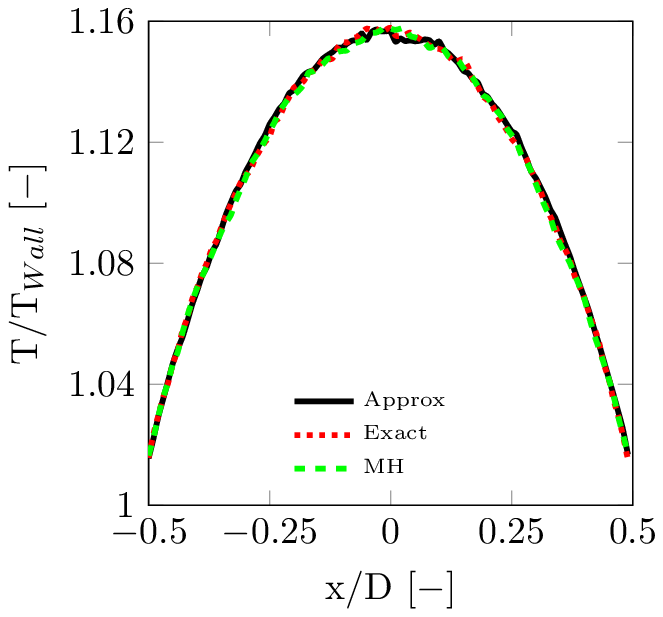}}\quad
\subfloat[Comparison of energy conservation methods.\label{fig:lowveloenergyc}]{\includegraphics{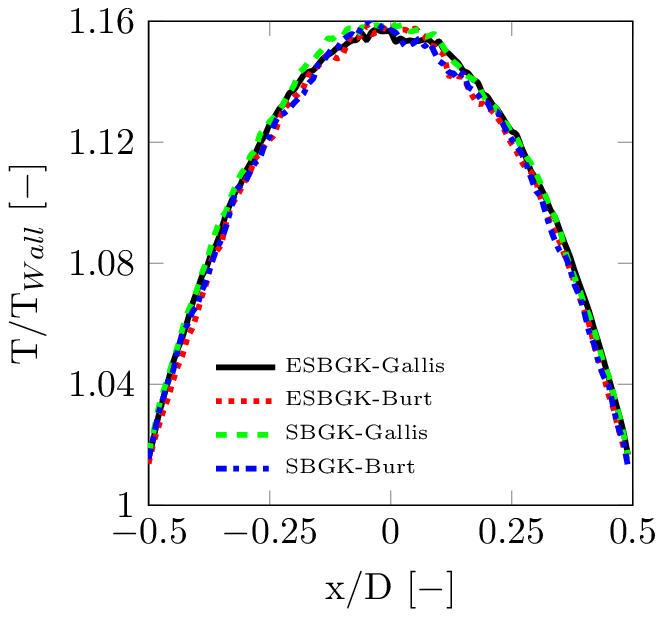}}\\
\caption{Simulation results for the low wall velocity case ($Kn\approx 0.014$, $v_{Wall}=\pm 250 \,\mathrm{m/s}$): (a) temperature profiles and (b) velocity profiles using different BGK models. (c) Comparison of ESBGK sampling methods.
(d) Comparison of energy conservation methods.}
\end{figure}

In the low velocity case, the plate velocity is $v_{Wall}=\pm 250 \,\mathrm{m/s}$ and the particle density is $n=1.29438\cdot 10^{20}\,\mathrm{1/m^3}$. 
This leads to a Knudsen number of $Kn\approx 0.014$.
The results of the temperature and velocity profiles between the plates for the different methods are shown in 
Fig. \ref{fig:lowvelo} and \ref{fig:lowvelob}.
The ESBGK as well as the SBGK method match the DSMC results very well whereas a clear difference is visible in the temperature plot for the BGK method.

Additionally, the three different methods of sampling from the ESBGK distribution (see Sec. \ref{sec:esbgkimpl}) are compared. Fig. \ref{fig:lowveloc}
shows that there is no visible difference in the temperature plot. 
Furthermore, the energy conservation schemes described in Sec. \ref{sec:encon} are compared in Fig. \ref{fig:lowveloenergyc}. Here again no significant difference
between the energy conservation schemes can be found. Therefore all other Couette flow simulations are done using only the energy conservation scheme of Gallis and Torczynski \cite{gallis2011investigation}.
Nevertheless, it is shown
later that the energy conservation scheme becomes significant in reproducing the correct heat flux values in hypersonic flows including shocks.
Finally, the computational time of the different methods is compared in Table \ref{tab:cputime}, showing that
there is no significant difference using the different models. 
\begin{table}
  \begin{center}
\def~{\hphantom{0}}
  \begin{tabular}{p{3cm}| C{1.5cm} C{1.5cm} C{1.5cm} C{1.5cm} C{1.5cm} C{1.5cm} C{2cm}}
  Model & ESBGK Appr. & ESBGK Exact & ESBGK MH & SBGK MH & SBGK AR & BGK$^*$& UBGK \linebreak MH $C_{ES}=0.1$\\ 
  \hline
  CPU time [s] per 1000 iterations & 78 & 78 & 83 & 83 & 82 & 75 & 94
  \end{tabular}
  \caption{\label{tab:cputime}CPU time of different models all simulated on one core. $^*$ The BGK method does not need a special sampling since only normal distributed random numbers are necessary.
    The normal distributed random numbers for all simulations are generated using the Ziggurat algorithm \cite{marsaglia2000ziggurat}.}
  \end{center}
\end{table}

\subsubsection{High Wall Velocity Case - $Kn\approx 0.014$}
\label{sec:highwallone}
\begin{figure}
\centering
\subfloat[Temperature\label{fig:highveloKna}]{\includegraphics{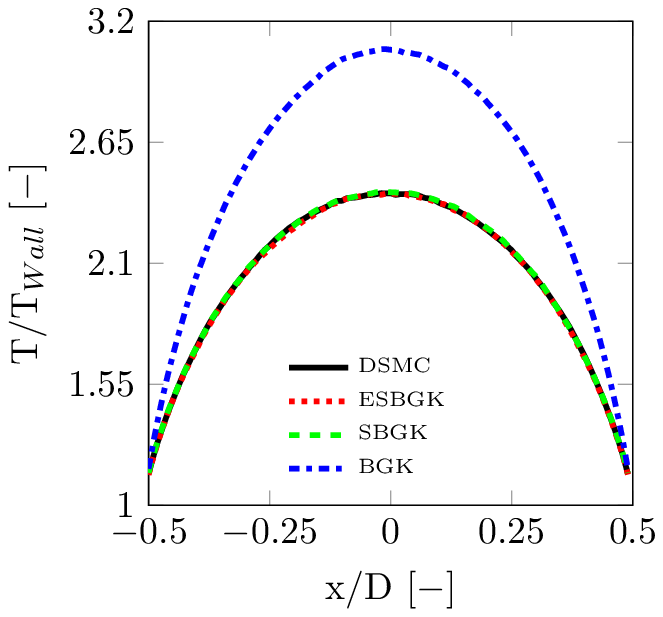}}\quad
\subfloat[Velocity\label{fig:highveloKnb}]{\includegraphics{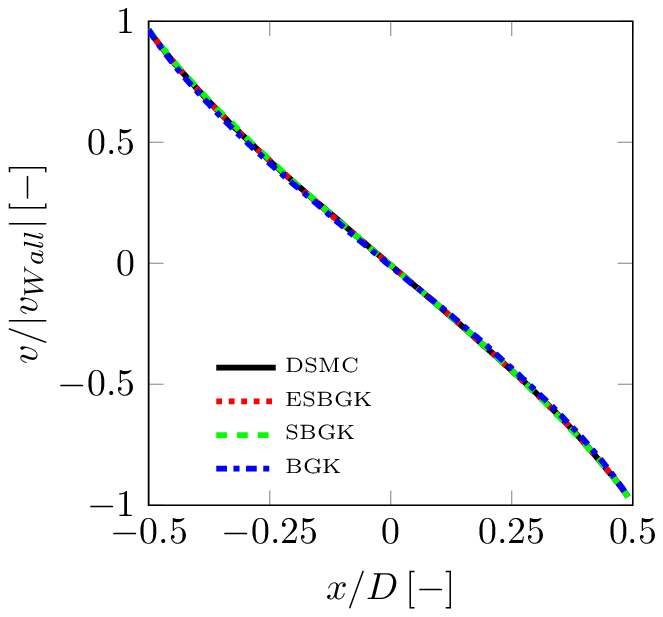}}\\
\subfloat[Temperature\label{fig:highveloKnc}]{\includegraphics{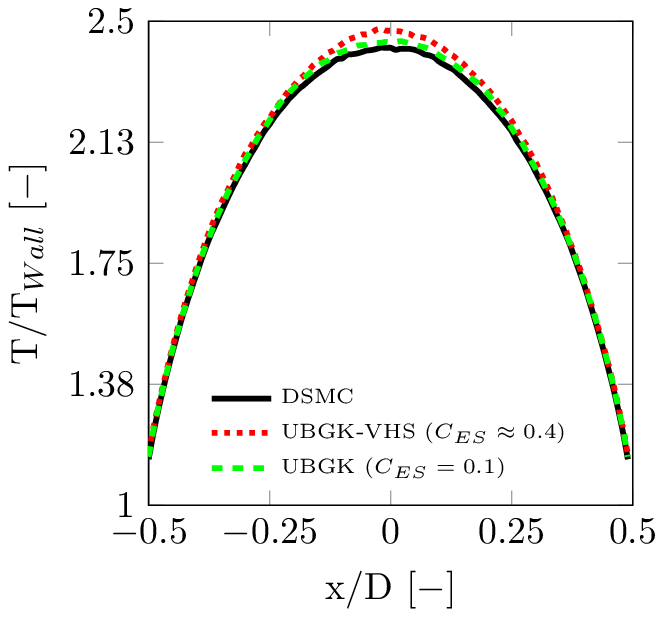}}\quad
\subfloat[Velocity\label{fig:highveloKnd}]{\includegraphics{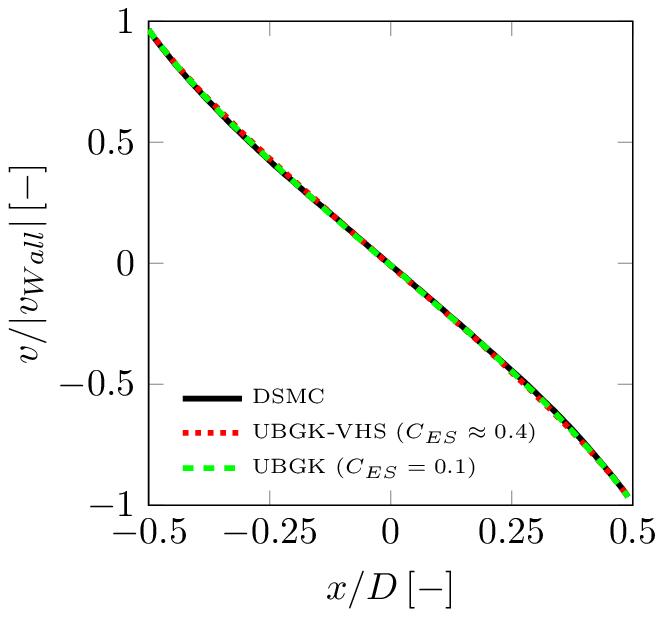}}
\caption{(a) \& (b): Velocity and temperature profiles of high wall velocity Couette flow ($Kn\approx 0.014$) using different BGK models. 
(c) \& (d): Velocity and temperature profiles of high wall velocity Couette flow ($Kn\approx 0.014$) using the UBGK method with different $C_{ES}$.}
\end{figure}

The high velocity case has a plate velocity of $v_{Wall}=\pm 750 \,\mathrm{m/s}$ and a particle density of $n=1.29438\cdot 10^{20}\,\mathrm{1/m^3}$.
So, the Knudsen number is again $Kn\approx 0.014$. The results of this case for the temperature and velocity profiles between the plates are shown in 
Fig. \ref{fig:highveloKna} and \ref{fig:highveloKnb}.
As in the case of the low wall velocity, the ESBGK as well as the SBGK match the DSMC result very well whereas the BGK model overestimates the temperature. 
Furthermore, the UBGK model was tested with this case. The results compared with the DSMC method are shown in Fig. \ref{fig:highveloKnc} and \ref{fig:highveloKnd}.
The temperature overestimates the DSMC result for the VHS equilibrium solution of $C_{ES}\approx 0.4$ (Eq. \eqref{eq:cesvhs}). For decreasing $C_{ES}$, the UBGK 
temperature approaches the DSMC temperature. Clearly, the UBGK method would match the DSMC solution for $C_{ES}=0$ where the UBGK model reduces to the SBGK model.
Nevertheless, this shows the disadvantage of the UBGK method: it is not clear how to chose the appropriate $C_{ES}$.

\subsubsection{High Wall Velocity Case - $Kn\approx 0.14$}
\begin{figure}
\centering
\subfloat[Temperature\label{fig:highveloKn01a}]{\includegraphics{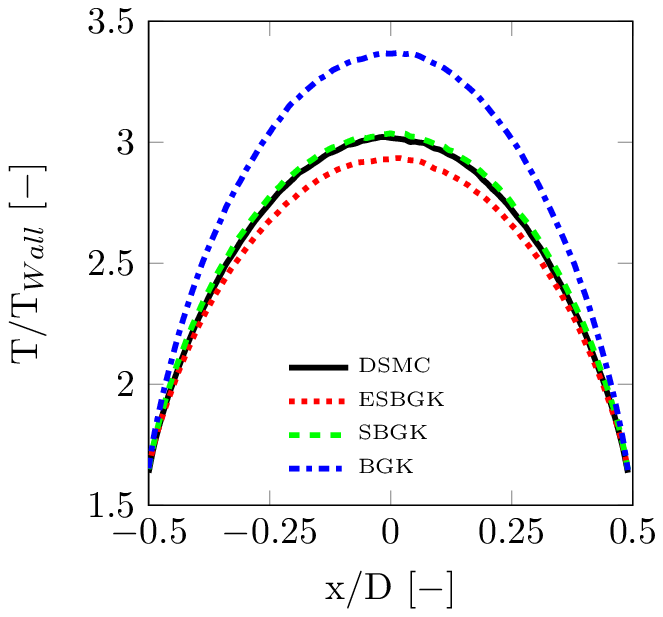}}\quad
\subfloat[Velocity\label{fig:highveloKn01b}]{\includegraphics{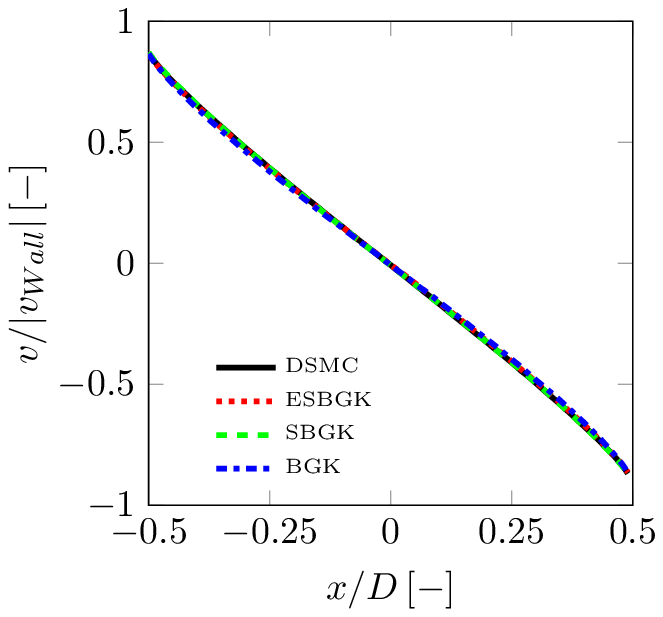}}\\
\subfloat[Temperature\label{fig:highveloKn01c}]{\includegraphics{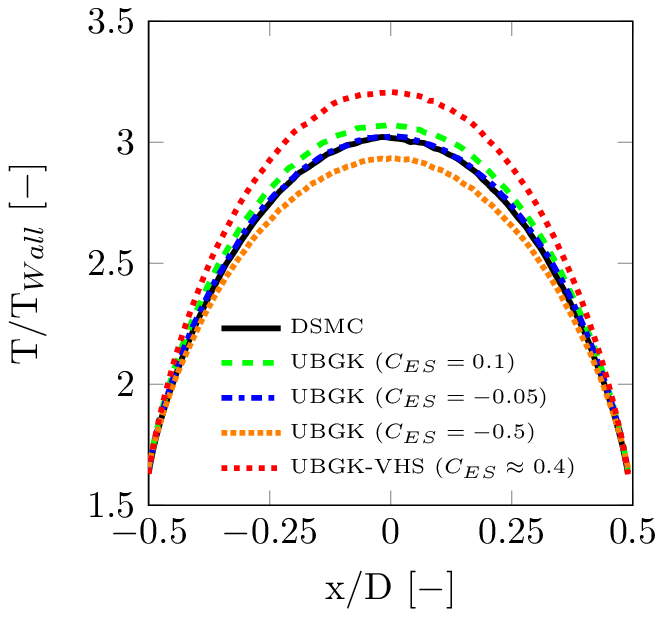}}\quad
\subfloat[Velocity\label{fig:highveloKn01d}]{\includegraphics{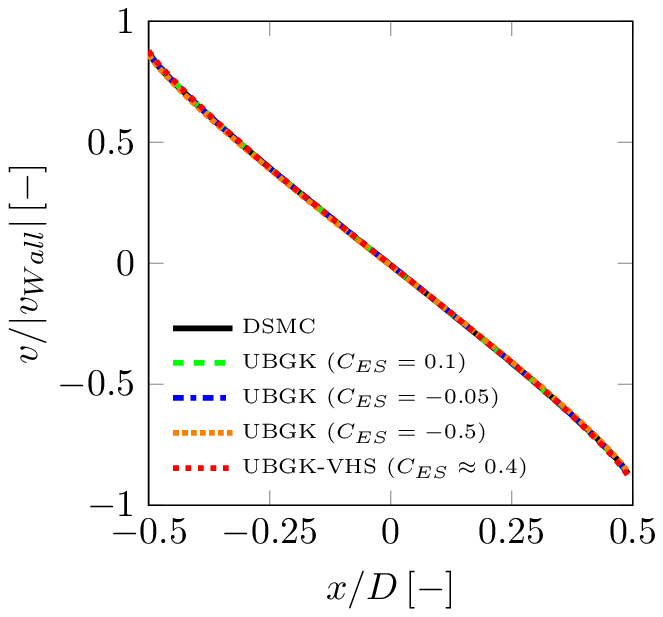}}
\caption{(a) \& (b): Velocity and temperature profiles of high wall velocity Couette flow ($Kn\approx 0.14$) using different BGK models. 
(c) \& (d): Velocity and temperature profiles of high wall velocity Couette flow ($Kn\approx 0.14$) using the UBGK method with different $C_{ES}$.}
\end{figure}

The conditions are the same as described in Sec. \ref{sec:highwallone}. The only difference is the particle density of $n=1.29438\cdot 10^{19}\,\mathrm{1/m^3}$,
which leads to a Knudsen number of $Kn\approx 0.14$. Therefore, this case is not in the continuum regime. The temperature and velocity profiles between the plates for the 
different methods are shown in 
Fig. \ref{fig:highveloKn01a} and \ref{fig:highveloKn01b}.
Here, the SBGK method shows good agreement with the DSMC results, whereas the BGK method overestimates and the ESBGK method underestimates the resulting temperature.
Fig. \ref{fig:highveloKn01c} and \ref{fig:highveloKn01d} show additionally an advantage of the UBGK method: the UBGK results can be tuned  
to match the DSMC results with $C_{ES}=-0.05$ in the non-continuum regime.

\subsubsection{High Wall Velocity Case - $Kn\approx 1.4$}
\begin{figure}
\centering
\subfloat[Temperature\label{fig:highveloKn1.4a}]{\includegraphics{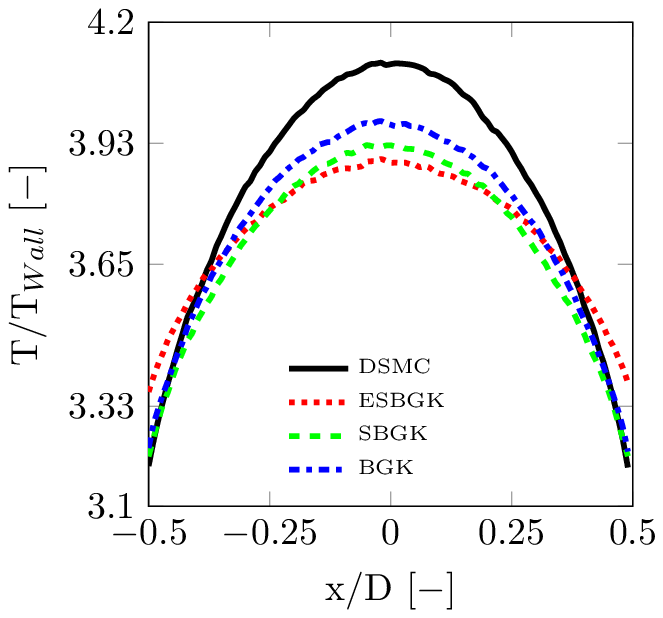}}\quad
\subfloat[Velocity\label{fig:highveloKn1.4b}]{\includegraphics{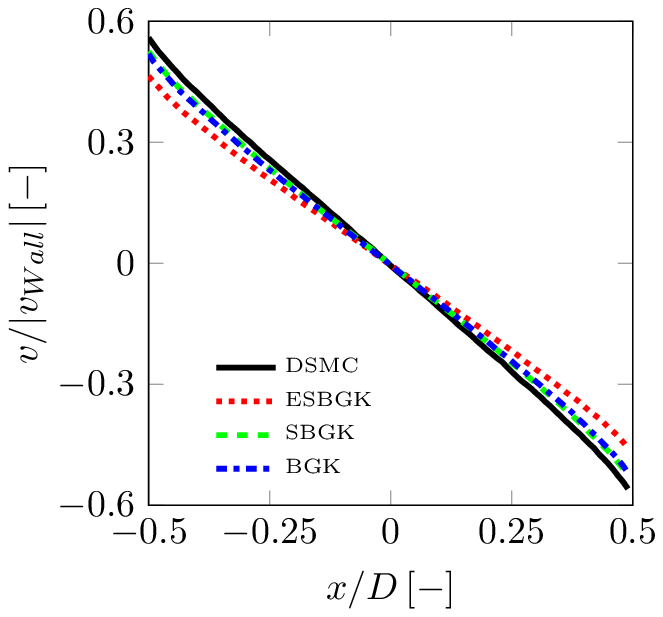}}\\
\subfloat[Temperature\label{fig:highveloKn1.4c}]{\includegraphics{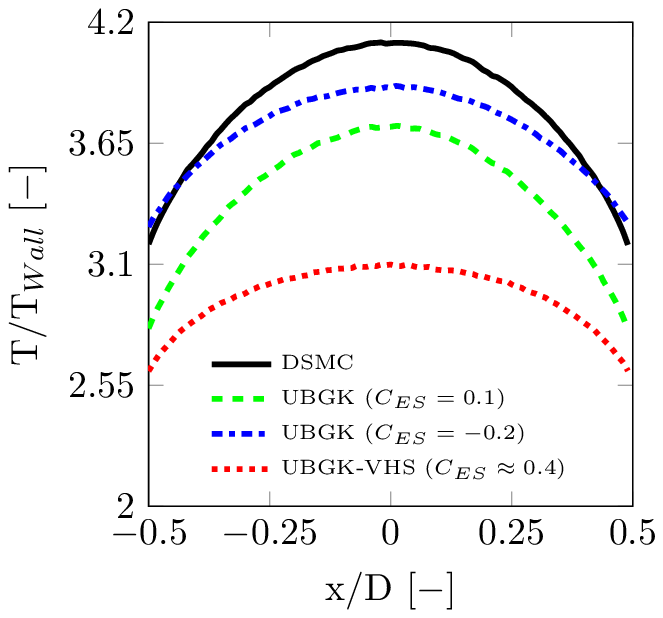}}\quad
\subfloat[Velocity\label{fig:highveloKn1.4d}]{\includegraphics{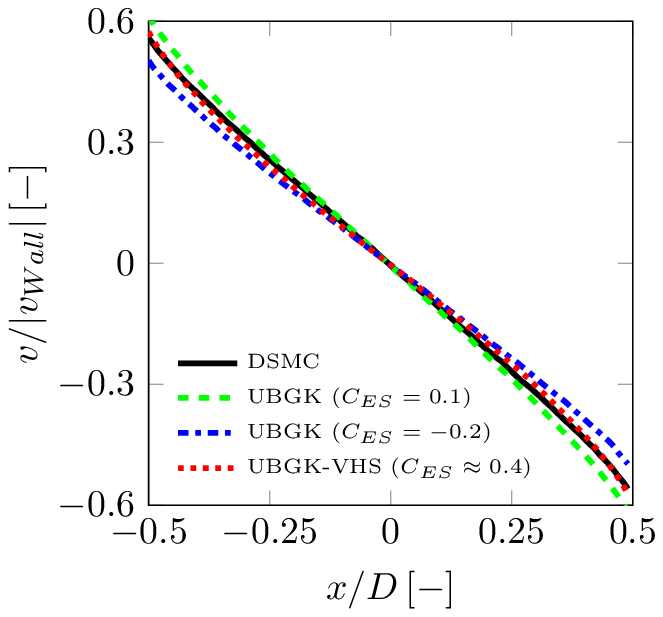}}
\caption{(a) \& (b): Velocity and temperature profiles of high wall velocity Couette flow ($Kn\approx 1.4$) using different BGK models. 
(c) \& (d): Velocity and temperature profiles of high wall velocity Couette flow ($Kn\approx 1.4$) using the UBGK method with different $C_{ES}$.}
\end{figure}

Finally, the Knudsen number is changed to $Kn\approx 1.4$. In such a rarefied condition, none of the BGK methods is able to reproduce 
the DSMC results as shown in Fig. \ref{fig:highveloKn1.4a}-\ref{fig:highveloKn1.4d}. 
Even the UBGK method is not capable to reproduce the correct values with the constrained $C_{ES}\in [-0.5,1)$.

\subsection{70$^\circ$ Blunted Cone}
The 70$^\circ$ blunted cone described in Allegre et al.\cite{Allegre1997} is chosen to compare the different BGK models with DSMC. The UBGK model was not used
in this investigation. As discussed in Sec. \ref{sec:ubgk} it is not clear how to choose $C_{ES}$ for such complex applications.
All simulations were carried out for Argon with the inflow conditions of Table \ref{tab:70cone}. 
The angle of attack is set to $\alpha=0^\circ$. The simulations were performed in 3D to compare the 3D performance between the BGK models
and DSMC. All simulations (DSMC as well as all BGK) are performed on the same coarse simulation mesh with 325200 cells. As described in paragraph \ref{sec:couette}, 
an octree method is used in DSMC to resolve the mean free path \cite[]{Pfeiffer2013}. The same octree method is used to subdivide the cells for the BGK methods. Only a certain number
of particles per cell form a subcell in the BGK methods to increase the spatial resolution.
The points \{A,B,C,D\} in Fig. \ref{fig:shockdistancea} correspond to the points in the following line plots.

\begin{table}
  \begin{center}
\def~{\hphantom{0}}
  \begin{tabular}{c | C{2cm} C{2cm} C{2cm} C{2cm}}
   &$v_\infty$ $\left[\mathrm{ms^{-1}}\right]$ &  $T_\infty$ $[\mathrm{K}]$ &  $n_\infty$ $[\mathrm{m^{-3}}]$ & $Kn_\infty$ \\
  \hline
  Set 1 & 1502.57 & 13.3 & $3.715\cdot10^{20}$ & $\approx 0.033$\\
  Set 2 & 1502.57 & 13.3 & $1.115\cdot10^{21}$ & $\approx 0.011$
  \end{tabular}
  \caption{\label{tab:70cone}Inflow conditions of 70$^\circ$ cone test case.}
  \end{center}
\end{table}

\subsubsection{Set 1}
Set 1 is a rarefied test case with a Knudsen number of $Kn\approx 0.033$. To resolve the mean free path and the collision frequency in the 3D DSMC simulation a particle number 
of $N_{DSMC}=2.2\cdot 10^7$ and a time step of $\Delta t_{DSMC}=1\cdot 10^{-7}\,\mathrm{s}$ are necessary. 

The BGK method has similar requirements
as the CFD method. 
The time step can be found using a classic Courant-Friedrichs-Lewy condition with the stream velocity and the speed of sound as described in Mirza et al.\cite{MIRZA2017269}.
To resolve the temperature and velocity gradients a certain number of cells is required. Additionally, a certain number of 
particles per cell is required to represent the moments of the distribution function. 

\begin{figure}
\centering
\subfloat[Heat flux]{\includegraphics{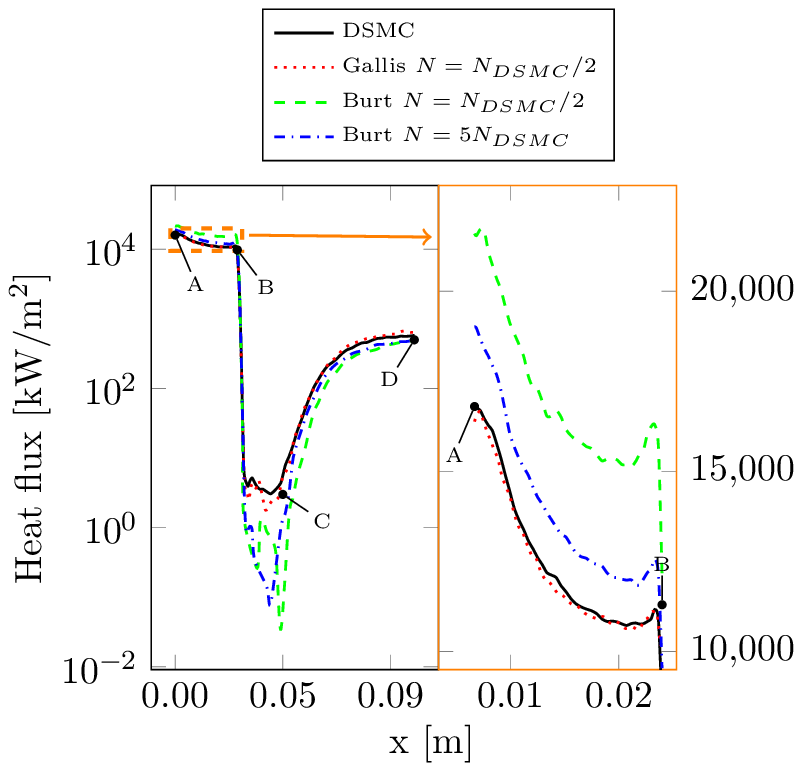}}
\subfloat[Temperature over stagnation stream line.]{\includegraphics{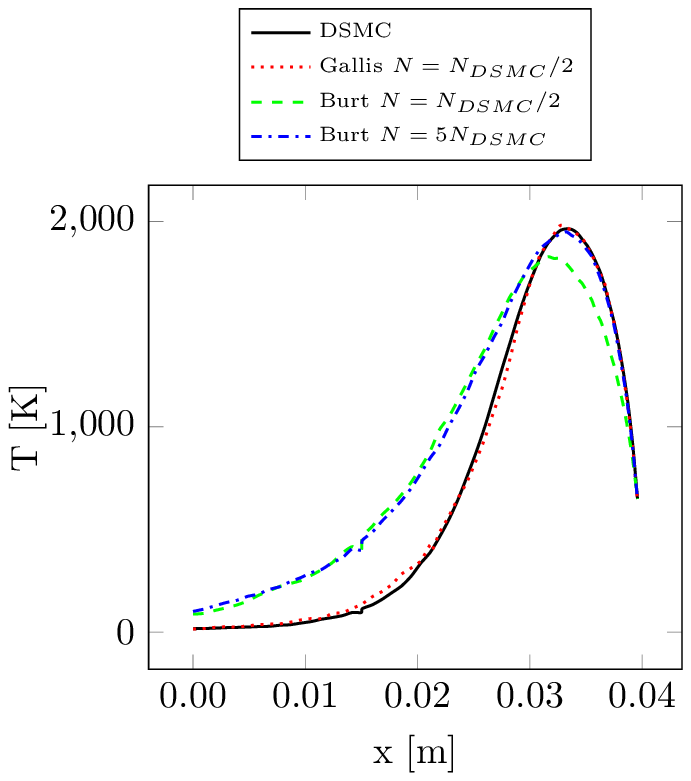}}
\caption{\label{fig:esbgkset1comp}Cone Set 1 using ESBGK with different energy conservation methods. a: Heatflux plot with magnification of the shield front region. 
The points \{A,B,C,D\} correspond to the points depicted in Fig. \ref{fig:shockdistancea}.
b: Comparison of the temperature (shock structure) plot over the stagnation stream line between different models.
Particle number and time step for DSMC: $N_{DSMC}=2.2\cdot 10^7$,
$\Delta t_{DSMC}=1\cdot 10^{-7}\,\mathrm{s}$. Time step for the ESBGK model: $\Delta t=\Delta t_{DSMC}$.}
\end{figure}

Here, the energy conservation scheme becomes crucial. While the energy conservation scheme of Gallis and Torczynski \cite{gallis2011investigation} 
shows good results with at least 7 to 10 particles per cell, the
energy conservation of Burt and Boyd \cite{burt2006evaluation} requires significantly more particles per cell. The comparison of the ESBGK method with the different energy conserving schemes is shown in 
Fig. \ref{fig:esbgkset1comp}. 

While the scheme of Gallis compares well with DSMC for the shock profile as well as heat flux on the surface with half the particles, the Burt scheme requires as many as five times more particles than DSMC to produce similar results. It is expected that the shock structure as well as the heat flux will also match the DSMC results with the Burt scheme if the particle number is further increased.
However, the particle number was not further increased due to the computational effort and the fact that a similar behavior was already shown in Tumuklu et al.\cite{tumuklu2016particle}.
Therefore, it is recommended to use the Gallis scheme for the ESBGK case as is done in the following simulation. 

\begin{figure}
\centering
\subfloat[Heat flux]{\includegraphics{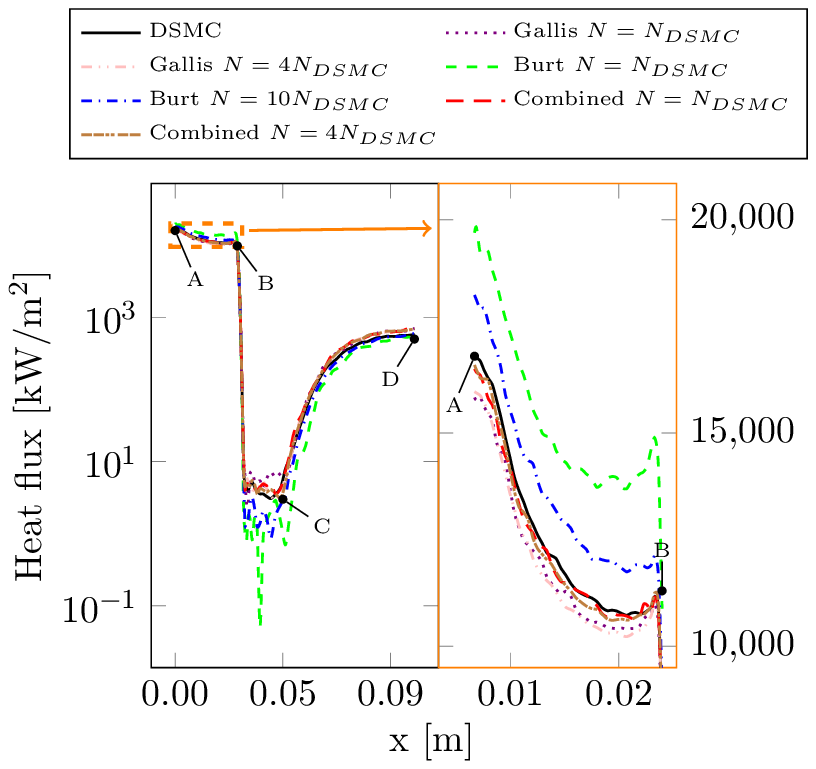}}
\subfloat[Temperature over stagnation stream line.]{\includegraphics{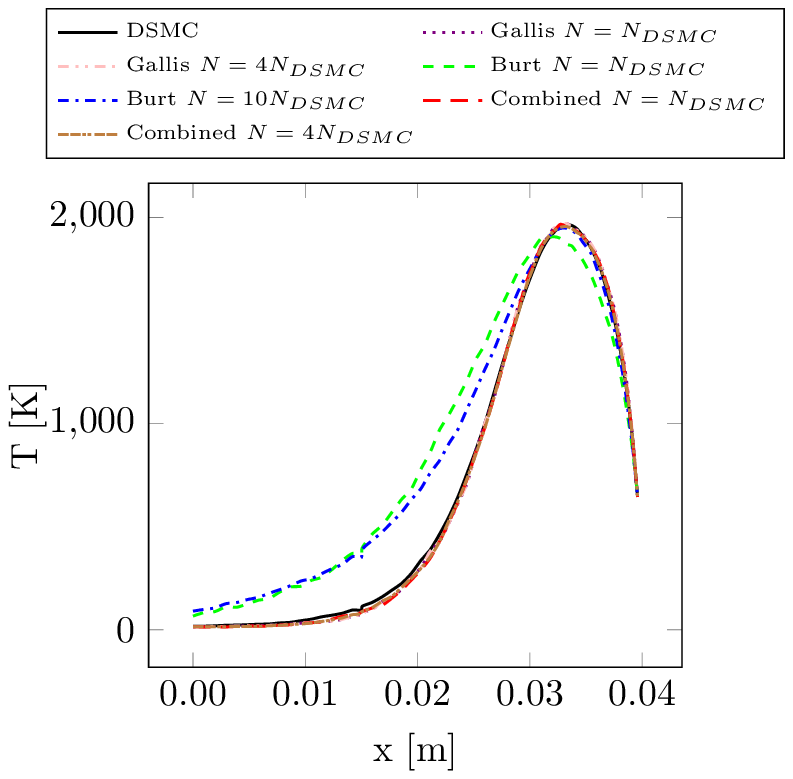}}
\caption{\label{fig:sbgkset1comp}Cone Set 1 using SBGK with different energy conservation methods. a: Heatflux plot with magnification of the shield front region. 
The points \{A,B,C,D\} correspond to the points depicted in Fig. \ref{fig:shockdistancea}.
b: Comparison of the temperature (shock structure) plot over the stagnation stream line between different models.
Particle number and time step for DSMC: $N_{DSMC}=2.2\cdot 10^7$,
$\Delta t_{DSMC}=1\cdot 10^{-7}\,\mathrm{s}$. Time step for the ESBGK model: $\Delta t=\Delta t_{DSMC}$.}
\end{figure}

The interpretation of the results for the energy conservation in the SBGK case is less obvious. As shown in Fig. \ref{fig:sbgkset1comp}, the Gallis scheme matches the shock profile very well and needs much less particles than the 
Burt scheme to converge. However, the Gallis scheme underestimates the heat flux on the surface. One possible reason for this behavior and the difference to the ESBGK scheme is
that the SBGK model produces skewed distribution functions due to the dependence on the third central moment (the heat flux vector). The energy conservation scheme of Gallis uses the whole distribution
function including non-relaxing particles, which seems to blur the skeweness of the distribution function too much, resulting in an incorrect heat flux vector. The ESBGK model seems to be more robust 
due to the fact that it produces non-skewed distribution functions that are symmetric in each direction. 

To overcome the problem of the SBGK model, while using moderate particle numbers, a combination of both energy conserving algorithms can be used. Test simulations have shown that the main 
problem of the Burt approach are cases with one or two relaxing particles per cell. Therefore, it is proposed to use the Gallis scheme if less than three relaxing particles are in the cell
and use the Burt scheme otherwise. All further SBGK simulations are done with this combined energy conservation scheme.

Due to these requirements, the required particle numbers can only be halved in the BGK simulations compared with DSMC for this case.

\begin{figure}
\centering
\subfloat[Heat flux\label{fig:set1heatforcea}]{\includegraphics{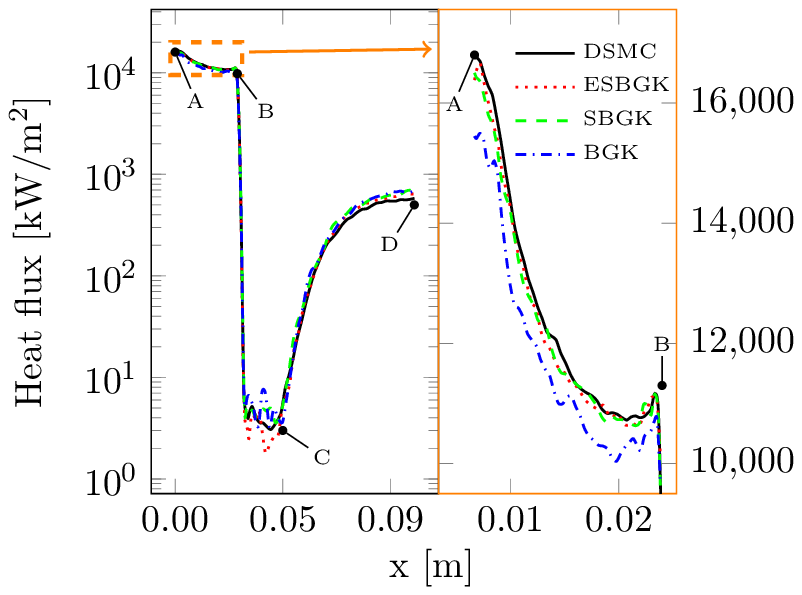}}
\subfloat[Pressure x direction\label{fig:set1heatforceb}]{\includegraphics{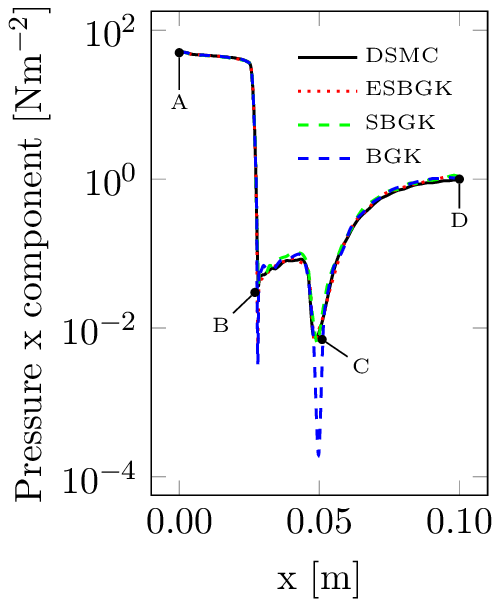}}\\
\centering{\subfloat[Temperature over stagnation stream line.\label{fig:set1heatforcec}]{\includegraphics{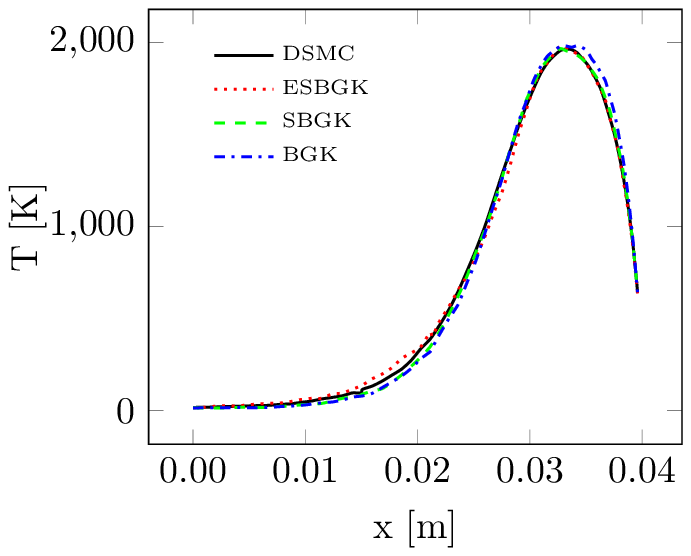}}}
\caption{Cone Set 1. a: Heatflux plot with magnification of the shield front region. b: Pressure on shield in x direction.
The points \{A,B,C,D\} correspond to the points depicted in Fig. \ref{fig:shockdistancea}.
c: Comparison of the temperature (shock structure) plot over the stagnation stream line between different models.
Particle number and time step for DSMC: $N_{DSMC}=2.2\cdot 10^7$,
$\Delta t_{DSMC}=1\cdot 10^{-7}\,\mathrm{s}$. Particle number and time step for the BGK models: $N=N_{DSMC}/2$, $\Delta t=\Delta t_{DSMC}$.}
\end{figure}

The comparison of the heat flux and pressure in x-direction between the different methods is shown in Fig. \ref{fig:set1heatforcea} and \ref{fig:set1heatforceb}. 
The SBGK with the combined energy conservation scheme as well as the ESBGK model with the Gallis scheme match the DSMC results very well, 
whereas the standard BGK model with the Gallis scheme underestimates the heat flux.
Fig. \ref{fig:set1heatforcec} shows the temperature plot over the stagnation stream line. 
The SBGK model agrees with the DSMC results slightly better than the ESBGK model. However, the predicted early onset of the 
temperature increase using the ESBGK model resulting in a wider temperature profile is less pronounced as expected.

\begin{figure}
\centering
\subfloat[Heat flux\label{fig:shockdistancedetailSet1a}]{\includegraphics{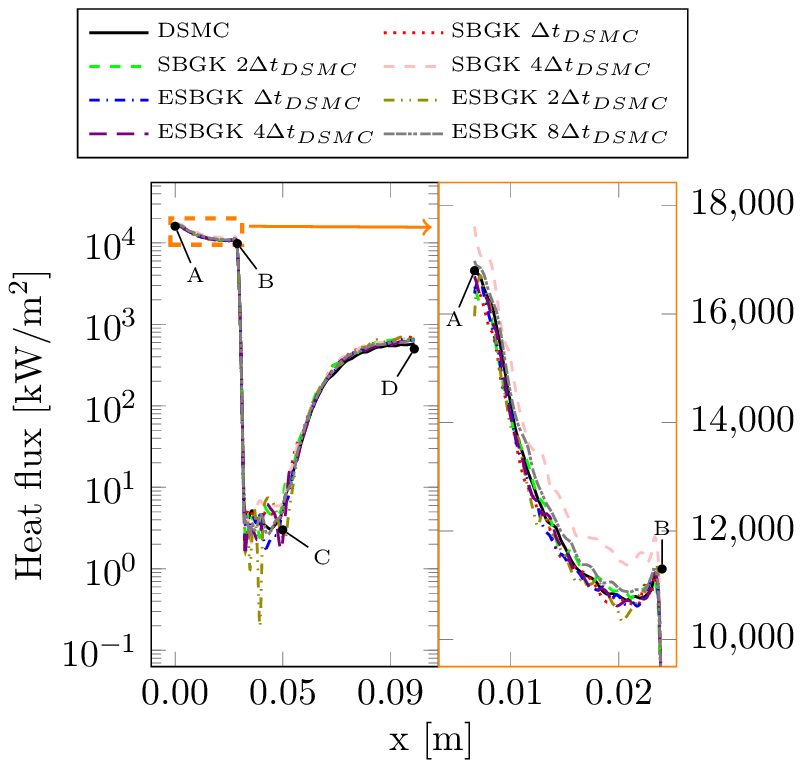}}
\subfloat[Temperature over stagnation stream line.\label{fig:shockdistancedetailSet1b}]{\includegraphics{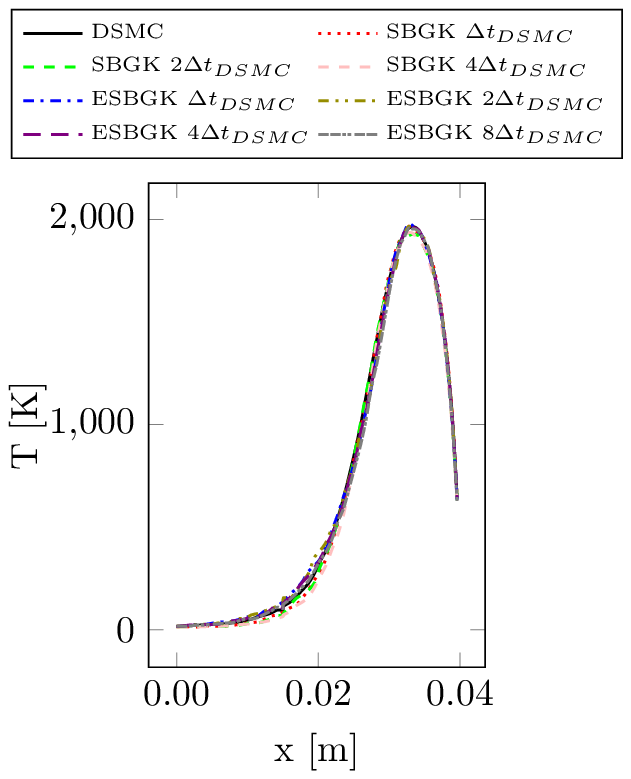}}
\caption{Comparison of the heat flux and temperature plot over the stagnation stream line between different time step discretizations
using ESBGK and SBGK.}
\end{figure}

Furthermore, a time step discretization study was conducted for the SBGK and ESBGK model. 
The results concerning the heat flux and the shock structure of this investigation 
are shown in Fig. \ref{fig:shockdistancedetailSet1a} and \ref{fig:shockdistancedetailSet1b}.
Here, the ESBGK model allows greater time steps, probably due to the fact that the same energy conservation method is used independent from the number of relaxing particles in the cell.
The heat flux and the shock structure using the ESBGK method match the DSMC results
well up to $\Delta t=4\Delta t_{DSMC}$, whereas the maximum time step for the SBGK method is $\Delta t=2\Delta t_{DSMC}$.

A comparison of the computational time is shown in Table \ref{tab:cputimeset1}. 
The coarsest discretization with acceptable results in terms of the heat flux and shock structure reduces the CPU time by a factor of $\approx 2.5$ using the SBGK method and 
by a factor of $\approx 4.8$ using the ESBGK method to reach the same simulation time.
\begin{table}
  \begin{center}
\def~{\hphantom{0}}
  \begin{tabular}{l | C{2cm} C{2cm} C{3cm} C{4cm}}
   & Particle Number $N$ &  Time step $\Delta t$ [s] &  CPU Time / 300 iterations [s] & CPU Time / $3\cdot 10^{-5}\,\mathrm{s}$ Simulation time [s] \\
  \hline

  DSMC & $2.2\cdot 10^7$ & $1\cdot 10^{-7}\,\mathrm{s}$  & 230 & 230 \\
  ESBGK & $N_{DSMC}/2$ & $\Delta t_{DSMC}$ & 152 & 152\\
  SBGK & $N_{DSMC}/2$ & $\Delta t_{DSMC}$ & 175 & 175\\
  BGK & $N_{DSMC}/2$ & $\Delta t_{DSMC}$ & 154 & 154\\
  SBGK & $N_{DSMC}/2$ & $2\Delta t_{DSMC}$ & 190 & 95\\
  ESBGK & $N_{DSMC}/2$ & $4\Delta t_{DSMC}$ & 190 & 47.5
  \end{tabular}
  \caption{\label{tab:cputimeset1} Comparison of CPU time between the different methods for Set 1. The CPU time is the time per node with 24 cores on a Intel Xeon CPU E5-2680 v3.}
  \end{center}
\end{table}

\subsubsection{Set 2}
\begin{figure}
\centering{\subfloat[Heat flux \label{fig:set2heatforcea}]{\includegraphics{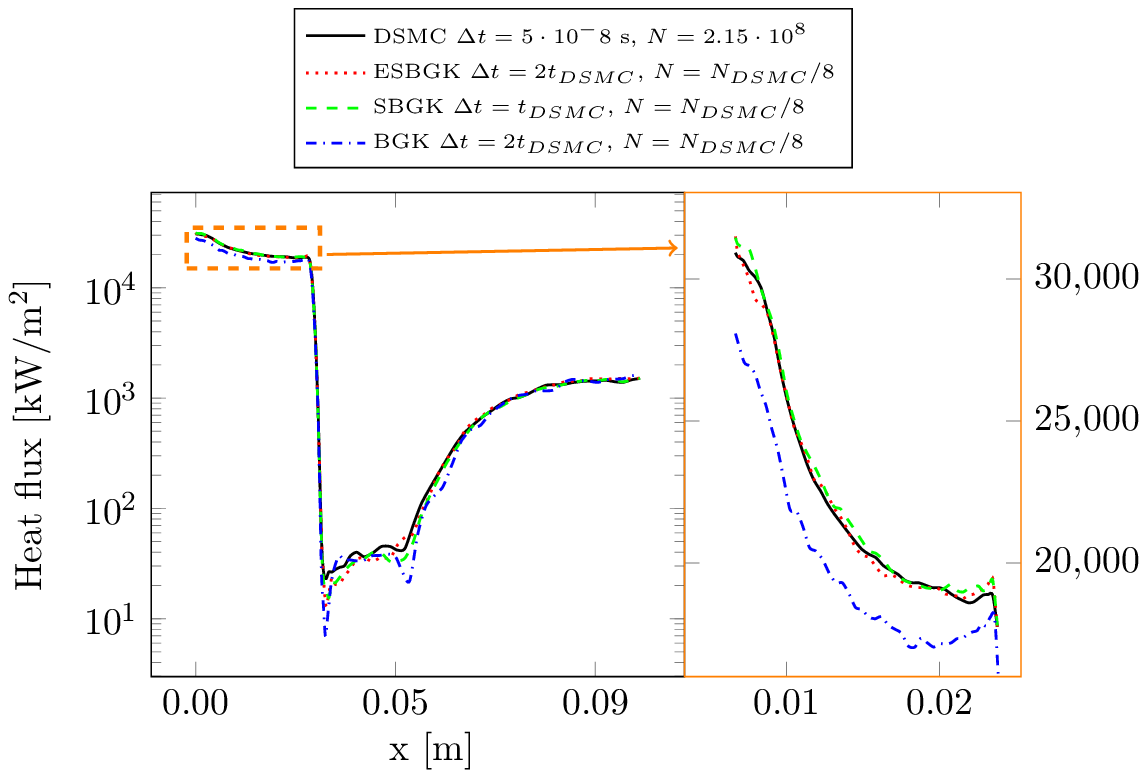}}}\\
\centering
\subfloat[Pressure x direction\label{fig:set2heatforceb}]{\includegraphics{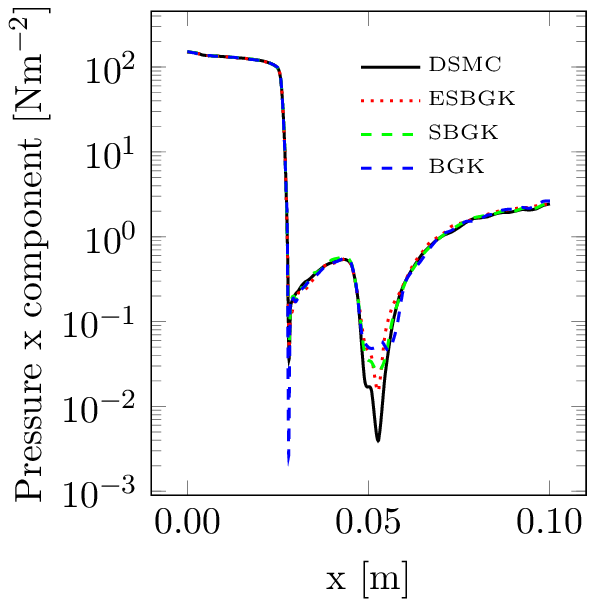}}\quad
\subfloat[Temperature over stagnation stream line.\label{fig:shockdistancedetaila}]{\includegraphics{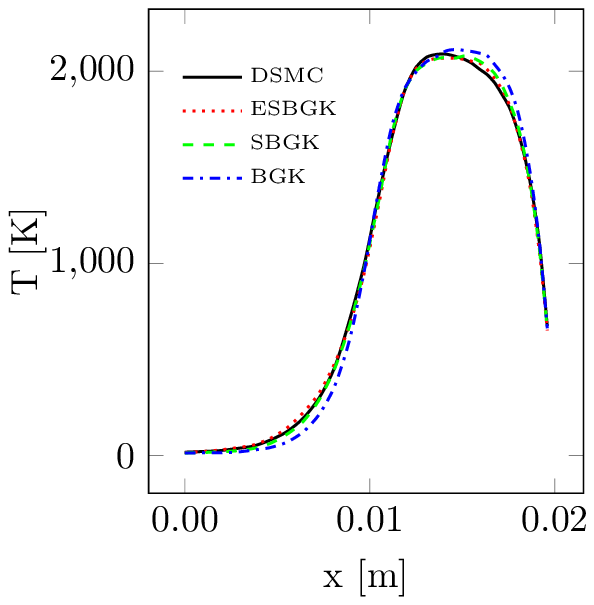}}
\caption{Comparison of DSMC with the different BGK models. Particle number and time step for DSMC: $N_{DSMC}=2.15\cdot 10^8$,
$\Delta t_{DSMC}=5\cdot 10^{-8}\,\mathrm{s}$. }
\end{figure}

\begin{figure}
\centering
\subfloat[DSMC $\leftrightarrow$ ESBGK\label{fig:shockdistancea}]{\includegraphics{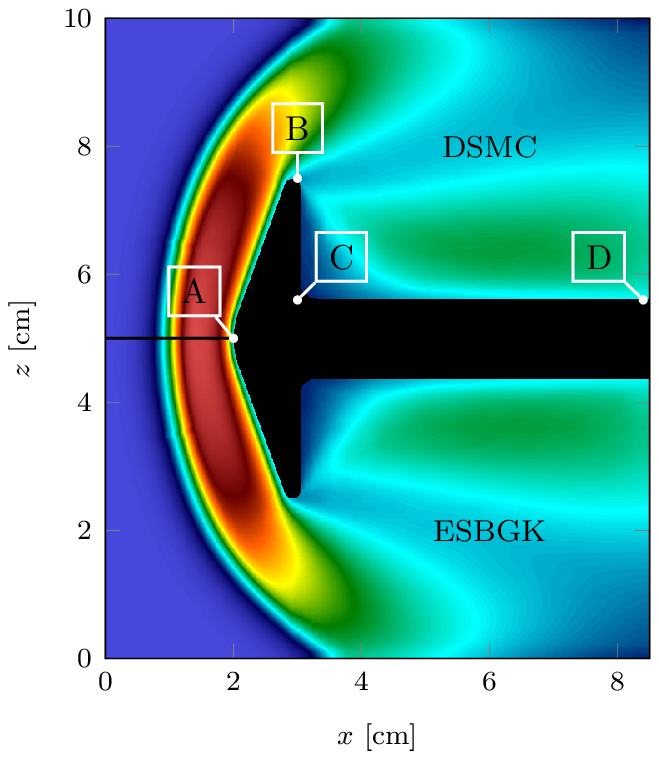}}\quad
\subfloat[DSMC $\leftrightarrow$ SBGK\label{fig:shockdistanceb}]{\includegraphics{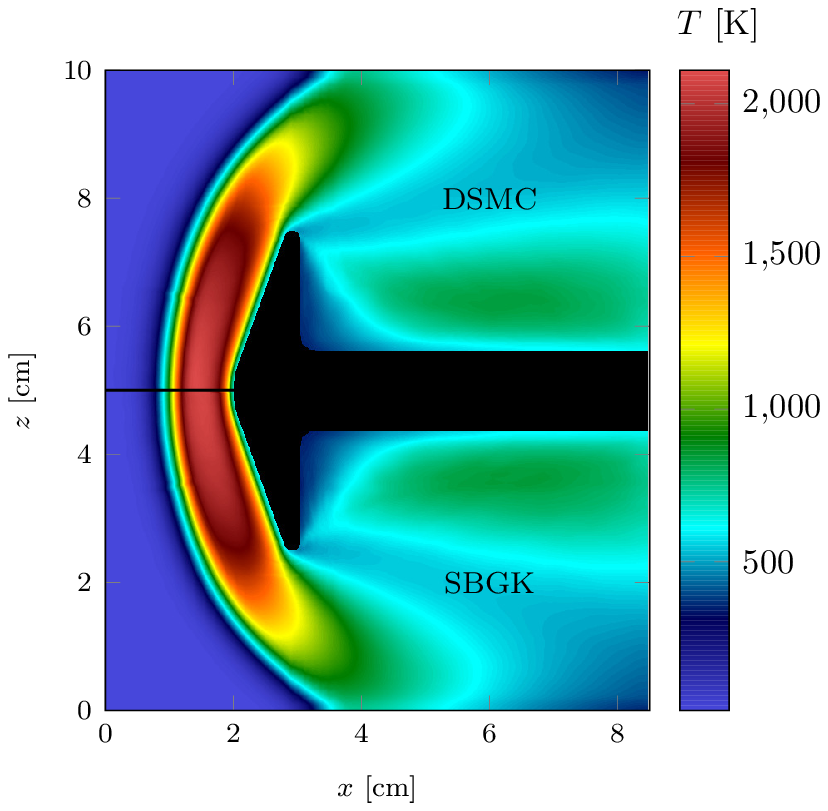}}
\caption{Temperature plots of ESBGK, SBGK and DSMC to compare the shock structure.}
\end{figure}

The conditions of Set 2 are given in Table \ref{tab:70cone}, where a higher density is assumed with a Knudsen number of $Kn\approx 0.011$.
To resolve the mean free path as well as the collision frequency in the 3D DSMC simulation
a particle number of $N_{DSMC}=2.15\cdot 10^8$ and a time step of $\Delta t_{DSMC}=5\cdot 10^{-8}\,\mathrm{s}$ are necessary.
The comparison of the heat flux and pressure in x-direction between the different methods is shown in Fig. \ref{fig:set2heatforcea} and
\ref{fig:set2heatforceb}. 
Here, the ESBGK and BGK method are using the energy conservation scheme of Gallis whereas the SBGK method uses the combined scheme.
Again, the SBGK as well as the ESBGK model match the DSMC results very well, whereas the standard BGK model underestimates the heat flux.
All simulations with the different BGK models are performed with eight times less particles $N=N_{DSMC}/8$. Furthermore, the
ESBGK and BGK model are using a two times larger time step $\Delta t=2\Delta t_{DSMC}$ whereas the SBGK model needs the same time step as in DSMC
$\Delta t=\Delta t_{DSMC}$.
In this case, the shock structure of the ESBGK and SBGK method match the DSMC results equally well
as shown in Fig. \ref{fig:shockdistancea} and \ref{fig:shockdistanceb} and in more detail in 
Fig. \ref{fig:shockdistancedetaila}.


A comparison of the CPU time is depicted in Table \ref{tab:cputimeset2}. In this case, the BGK methods can substantially save computational time
compared with DSMC. For the coarsest discretization with acceptable results in terms of the heat flux and shock structure (ESBGK: $N=N_{DSMC}/8$,
$\Delta t =2\Delta t_{DSMC}$; SBGK: $N=N_{DSMC}/8$,
$\Delta t =\Delta t_{DSMC}$) the ESBGK model reduces the CPU time by a factor of $\approx 17.5$ whereas the SBGK reduces the CPU time by a factor of 
$\approx 8.1$ to reach the same simulation time.

\begin{table}
  \begin{center}
\def~{\hphantom{0}}
  \begin{tabular}{l | C{2cm} C{2cm} C{3cm} C{4cm}}
   & Particle Number $N$ &  Time step $\Delta t$ [s]&  CPU Time / 100 iterations [s] & CPU Time / $1\cdot 10^{-5}\,\mathrm{s}$ Simulation time [s] \\
  DSMC & $2.15\cdot 10^8$ & $5\cdot 10^{-8}\,\mathrm{s}$  & 820  & 1640\\
  ESBGK & $N_{DSMC}/8$ & $2\Delta t_{DSMC}$ & 95 & 95\\
  SBGK & $N_{DSMC}/8$ & $\Delta t_{DSMC}$ & 101 & 202\\
  BGK & $N_{DSMC}/8$ & $2\Delta t_{DSMC}$ & 90 & 90
  \end{tabular}
  \caption{\label{tab:cputimeset2} Comparison of CPU time between the different methods for Set 2. . The CPU time is the time per node with 24 cores on a Intel Xeon CPU E5-2680 v3.}
  \end{center}
\end{table}

\section{Conclusion and Outlook}
Different ways to sample from an arbitrary velocity distribution function were investigated. Using these methods, 
it was possible to use the Shakhov BGK and the unified BGK models in 
an efficient way in the context of particle simulations. Furthermore, different energy conservation scheme for particle based BGK methods 
were investigated in detail. 


The described methods were validated using Couette flow test cases with different wall velocities and Knudsen numbers. It was shown that 
the SBGK model performs well up to $Kn=0.14$ whereas the ESBGK model shows distinct differences to the DSMC simulations in this 
Knudsen number regime. Furthermore, it was demonstrated that it is possible to capture rarefied flow phenomena using the UBGK model 
by adapting the additional free parameter of the UBGK model. Nevertheless, it is currently not clear how this additional free parameter
should be defined. Therefore, the UBGK model is not useful for practical applications at the moment.
Additionally, the computational time of the different BGK methods were compared. Here, it was shown that the investigated
sampling methods are not significantly slower than the established method to sample from the ESBGK target function 
and enables the sampling from the SBGK and UBGK target distribution in an efficient way.

Further on, the BGK, SBGK and ESBGK models were compared with DSMC simulations based on the hypersonic flow around a 70$^\circ$ blunted cone to 
evaluate the capabilities to capture shock waves.
Here, it was shown that the energy conservation scheme is especially crucial for the SBGK method. The ESBGK is more robust concerning the 
energy conservation and the time step size. Furthermore, the shock profile was not significantly better reproduced by the SBGK model compared
with the ESBGK model.
However, the heat flux of the SBGK as well as the ESBGK model match the DSMC result very well. It was shown that for a low density
case (Case 1), the SBKG model can save up to a factor of $\approx 2.5$ and the EBKG model can save up to a factor of $\approx 4.8$ CPU time compared with DSMC. With increasing density, the BGK models are able to save 
significant CPU time compared with DSMC. It was demonstrated that the ESBGK model can save up to a factor of $\approx 17.5$ 
and the SBGK model can save up to a factor of $\approx 8.1$ CPU time compared with DSMC
for a shock-containing flow problem with a Knudsen number on the order of $Kn\approx 0.011$.

This behavior is especially very interesting for gas flows which cover a wide range of Knudsen numbers including continuum and rarefied gas regions as in nozzle expansion flows,
where the coupling of the proposed BGK methods with DSMC is required in order to save computational time.
The fact that DSMC and the investigated methods are both cell local Monte-Carlo based particle methods, makes a coupling very simple without the typical problems
of hybrid CFD-DSMC methods.

Summarising, the ESBGK model should be preferred over the SBGK model in the presented particle based context. The ESBGK model is able to handle larger time steps than the SBGK model 
at least until a more robust energy conservation scheme is found for the SBGK model. Additionally, a general proof exists for the ESBGK model that it always fulfills the H-theorem and 
it is guaranteed that $f^{ES}$ is positive in each situation.

In the next steps, the described method will be extented to molecules including internal energies. A possible implementation for rotational energies is already given 
in Burt and Boyd\cite{burt2006evaluation},
which was also already extended to vibrational energy modes in Tumuklu et al.\cite{tumuklu2016particle}. 
Tumuklu et al.\cite{tumuklu2016particle} showed a similar behavior of the computational efficiency comparing
ESBGK and DSMC as described here: the ESBGK method becomes more efficient for an increasing collision frequency compared with DSMC. Therefore, we also expect a reduction of computational time using a BGK 
method compared to DSMC in small Knudsen number flows including internal energies.

\section*{Acknowledgments}

The author gratefully acknowledges the Deutsche Forschungsgemeinschaft (DFG) for funding this research within the
project ``Partikelverfahren mit Strahlungsl\"oser zur Simulation hochenthalper Nichtgleichgewichts-Plasmen'' (project number 93159129). 
The author also thanks the High Performance Computing Center Stuttgart (HLRS) for granting the computational time that has allowed the execution of
the presented simulations.
 
\bibliography{mybibfile}
\end{document}